\documentclass[%
twocolumn,
groupedaddress,
 amsmath,amssymb,
]{revtex4-2}

\usepackage{graphicx}
\usepackage{dcolumn}
\usepackage{bm}
\usepackage{hyperref}

\hypersetup{
    colorlinks=true,
    linkcolor=blue,
    filecolor=magenta,      
    urlcolor=cyan,
    pdftitle={Sharelatex Example},
    bookmarks=true,
    citecolor=blue
}

\usepackage{gensymb}
\usepackage{graphicx}
\usepackage[]{siunitx}
\usepackage[tight,FIGTOPCAP,nooneline, bf, normalsize]{subfigure}
\usepackage{pgfplots}
\pgfplotsset{tick label style={font=\footnotesize},
    every x tick label/.append style={font=\footnotesize},
    label style={font=\footnotesize},
    legend style={font=\scriptsize, legend style={row sep=-2.5pt}},
    tick style={line width=0.5pt},
    major tick length={0.05cm},
    minor tick length={0.03cm},
    every tick/.append style={color=black},
    every addplot/.append style={line width=10pt},
    compat=1.18,
    }

\pgfplotsset{
    compat=newest,
    /pgfplots/legend image code/.code={%
        \draw[mark repeat=2,mark phase=2,#1] 
            plot coordinates {
                (0cm,0cm) 
                (0.15cm,0cm)
                (0.3cm,0cm)
            };
    },
}
\usepackage{pgfplotstable}
    \pgfkeys{/pgf/number format/.cd,set thousands separator={}}

\definecolor{jet7_1}{RGB}{0,114,189}
\definecolor{jet7_2}{RGB}{216,83,25}
\definecolor{jet7_3}{RGB}{237,177,32}
\definecolor{jet7_4}{RGB}{126,47,142}
\definecolor{jet7_5}{RGB}{119,172,48}
\definecolor{jet7_6}{RGB}{77,190,238}
\definecolor{jet7_7}{RGB}{162,20,47}

\usetikzlibrary{tikzmark}

\begin{document}

\preprint{APS/123-QED}

\title{Exploring phase sensitivity and limit of detection near the critical coupling of metasurfaces and its phase singularity}

\author{Lotfi Berguiga$^1$}
\email{lotfi.berguiga@insa-lyon.fr}
\author{Th\'{e}o Girerd$^1$}%
\author{Xavier Letartre$^2$}%
\author{Taha Benyattou$^1$}%
\author{C\'{e}cile Jamois$^1$}%
\author{Fabien Mandorlo$^1$}%
\author{Lydie Ferrier$^1$}%
\email{lydie.ferrier@insa-lyon.fr}

\affiliation{$^1$INSA Lyon, Ecole Centrale de Lyon, CNRS, Université Claude Bernard Lyon 1, CPE Lyon, INL, UMR5270, 69621 Villeurbanne, France}%
\affiliation{$^2$Ecole Centrale de Lyon, INSA Lyon, CNRS, Université Claude Bernard Lyon 1, CPE Lyon, INL, UMR5270, 69130 Ecully, France}

\date{\today}

\begin{abstract}
It is commonly accepted that phase singularities in refractive index sensors can provide highly sensitive detection. To address this issue, we studied the phase sensitivity and the limit of detection of Tamm photonic crystals used as temperature sensors, taken here as a model system by exploring critical coupling and its associated phase singularity. To finely tune the optical Tamm mode to critical coupling, the top metal layer is periodically nanostructured and controlled, enabling the investigation of optical Tamm resonances around phase singularities. We use a highly stable common-path interferometry setup based on digital holography, which allows us to measure extremely high phase sensitivity even at very low light intensities, while also providing information about reflectivity in the complex domain. We experimentally validate an analytical model based on Temporal Coupled Mode Theory, which fully explains the phase sensitivity response and the limit of detection of such resonant photonic sensors. We demonstrate that, although approaching a phase singularity drastically enhances phase sensitivity, no improvement of the limit of detection is expected. Furthermore, phase singularity has no effect on the limit of detection, and therefore on the efficiency of such sensors for practical applications. This work provides a comprehensive description of phase sensors operating at critical coupling, and challenges the commonly accepted assumption regarding the role of phase singularities in sensing.
\end{abstract}


\maketitle


\section{Introduction}
There is a significant challenge in designing optical sensors that are highly sensitive, inexpensive, and easy to use. Such sensors have a wide range of applications, for instance in health and food analysis, in the medical field for the detection of pathologies (such as pathogens \cite{blevins2021roadmap} or cancer \cite{kaur2022recent,parvin2021differential}), and in environmental monitoring, for example in the detection of harmful gases like volatile organic compounds (benzene, toluene, carbon monoxide, etc.) \cite{gallego2008indoor,mondal2021exhaled,sethi2013clinical}.
Several approaches have been proposed, including Surface Plasmon Resonances (SPR) \cite{liedberg1995biosensing,liedberg1983surface,singh2016spr}, waveguided modes \cite{fabricius1992gas,hao2017graphene,schubert1997refractive}, and guided resonances in photonic crystals \cite{liu2019optofluidic,drayton2019guided,pitruzzello2018photonic,barth2020common}. Optical sensors exploit the spatial confinement of light and the strong dependence of optical resonances on the geometry or refractive index of the material and its surrounding environment. In this way, the electromagnetic field serves as a probe for the analyte.
Another strategy to achieve high sensitivity is to focus on the phase change occurring at resonance upon reflection or transmission of light incident on a photonic resonator.
It has been shown \cite{grigorenko1999phase,kabashin1997interferometer,kabashin2009phase,boriskina2018sensitive,liu2023} that when the reflected light amplitude approaches zero, the phase of the light undergoes a step-like jump. This sharp local phase variation is highly sensitive to the surrounding environment of the structure (refractive index, geometry, etc.), and this effect has been experimentally demonstrated mainly in SPR structures \cite{song2017dispersion,kravets2013singular,sreekanth2018biosensing}. Recent studies have proposed the realization of highly sensitive sensors by exploiting phase singularities. One of the main challenges in this approach is the design and fabrication of structures operating close to the critical point, i.e. the phase singularity.

In photonic structures containing metallic layers, critical coupling (i.e. $R \rightarrow 0$) can be achieved either by tuning material properties—for instance using phase-change materials \cite{cueff2021reconfigurable,sreekanth2018ge2sb2te5,piper2014total,berguiga2021ultimate}, porous materials \cite{kim2022single}, or 2D materials such as transition metal dichalcogenides \cite{ma2023excitons,ermolaev2022topological}—or by carefully adjusting the geometry of the optical structure (layer thicknesses, material patterning \cite{mkhitaryan2017tunable,yue2023high,piper2014total,berguiga2021ultimate}).

Several research groups have experimentally reported a correlation between phase sensitivity $S_{\phi}$ and the minimum reflectivity $R_{\text{min}}$ \cite{tsurimaki2018topological,sakotic2021topological,vasic2014enhanced,nikitin1999surface}: the lower the reflectivity, the higher the phase sensitivity. However, these studies have so far only shown this trend empirically, without providing a clear analytical relationship between $S_{\phi}$ and $R_{\text{min}}$. In our work, we experimentally demonstrated this relationship using optical Tamm structures.

Optical Tamm modes are located at the interface between a dielectric Bragg mirror and a thin metallic layer. These optical interface states have been widely exploited for sensing applications—including gas sensing \cite{auguie2014tamm,zaky2020refractive}, temperature sensing \cite{maji2017hybrid,tsurimaki2018topological}, and protein detection \cite{buzavaite2020hybrid}—because they offer several advantages: they can be excited at normal incidence, the electromagnetic field is strongly localized at the metal–dielectric interface, and the overlap between the analyte and the field can be maximized \cite{zaky2020refractive}. Furthermore, in optical Tamm structures, critical coupling can be reached by precisely adjusting the thickness of each layer \cite{tsurimaki2018topological,auguie2015critical}, thus enabling the realization of highly sensitive sensors \cite{tsurimaki2018topological}. However, controlling layer thicknesses at the nanometer scale during deposition remains a very challenging task. Moreover, fabricating a single sample with fixed and well-controlled thicknesses does not allow to experimentally probe multiple conditions that cross the critical coupling point within the same device.\\
It is usually assumed that sensors operating at phase singularities exhibit superior performance, since the phase sensitivity increases as the reflectivity approaches zero \cite{tsurimaki2018topological,vasic2014enhanced,nikitin1999surface,ermolaev2022topological,ma2023excitons,Wang2024}. However, phase-singularity-based sensing has often been inadequately addressed for several reasons. First, phase sensitivity has never been fully characterized within a complete theoretical framework. Second, neither the behavior of the phase singularity nor the approach to critical coupling has been systematically investigated in the extreme limit by tuning the relevant parameters. Finally, the limit of detection (LOD)—the key figure of merit for evaluating sensor performance—has never been quantified as a function of proximity to the phase singularity. While LOD is typically reported for a given transducer, its correlation with phase singularity effects remains unexplored.

To address these points, four major challenges were overcome.
The first step was to design and fabricate structures operating very close to critical coupling. In this work, we demonstrated that optical Tamm resonances can be tuned by patterning the top metallic layer into one-dimensional periodic gratings \cite{ferrier2019tamm,gubaydullin2017tamm}. This patterning enables fine adjustment of the resonance towards the critical point through the grating parameters, leading to extremely low reflectivity ($R_{\text{min}} \approx 10^{-6}$) and consequently to very sharp phase transitions. Thus, we fabricated Tamm plasmon structures and optimized the geometry of the nanostructures to achieve this regime.

The second step was to measure the reflected signal extremely close to critical coupling, i.e., for very low intensity values and highly abrupt phase variations. To this end, we developed an original phase interrogation method based on interferometry and holography, capable of simultaneously measuring both the amplitude and phase of the reflected light \cite{Girerd2024}. This method also allows us to explore phase singularities in both the wavelength domain and the wavevector domain (Fourier space).
The third step was to establish a theoretical model of critical coupling and its associated phase singularity, in order to clearly distinguish between intrinsic transducer effects and artifacts from the measurement apparatus. In this work, we employed an analytical model based on Coupled Mode Theory, initially developed for SPR \cite{berguiga2021ultimate}, and adapted it to our sensor architecture. This model reveals the influence of key parameters—such as quality factor, wavelength sensitivity, resonance wavelength, and reflectivity amplitude—on phase sensitivity, and provides a clear analytical relationship between $S_{\phi}$ and $R_{\text{min}}$. Measurements of phase sensitivity on structures with varying proximity to critical coupling experimentally confirmed this relationship: the lower the reflectivity, the higher the phase sensitivity, thereby demonstrating extremely high phase sensitivities in optical Tamm structures.
Finally, in the fourth step, we related phase noise to the more commonly used intensity noise. By combining the theoretical expression of phase sensitivity with the relation between phase and intensity noise, we established an analytical model of the LOD for phase-interrogation-based sensing. Our experiments revealed that phase noise increases drastically near singularities, effectively canceling out the apparent benefits of enhanced phase sensitivity.
We therefore show that the LOD is not determined by phase singularity itself, but instead depends solely on three parameters: the quality factor, the wavelength sensitivity $S_{\lambda}$, and the intensity noise $\sigma_R$.\\

The paper is organized as follows. In the first part, we present an analytical model for a Tamm multilayer structure that establishes the relationship between phase sensitivity and key physical quantities. We highlight the crucial role of critical coupling on phase sensitivity and explain how it can be tuned by tailoring the geometrical parameters of the Tamm structure, in particular the grating design.
The second part is devoted to experimental results, showing how critical coupling is achieved and how extremely low reflectivity values at resonance, as well as phase singularities, can be reached. We characterize the position of phase singularities in wavevector space and investigate them through temperature-sensing measurements on different Tamm structures using the interferometric setup. Finally, we analyze both the phase sensitivity and the limit of detection as the system approaches critical coupling.

\section{Optical Tamm mode for sensing}
\subsection{Analytical derivation of phase sensitivity}
To derive an explicit relationship between the phase sensitivity $S_{\phi}$ and the minimum reflectivity $\left|r_{min}\right|=\sqrt{R_{min}}$, we use the Temporal Coupled Mode Theory (TCMT) formalism. The system is modeled as a single-mode resonator supporting a resonance at frequency $\omega_{res}$ (\si{\radian\per\second}), with internal losses (i.e., absorption) and radiative losses characterized by the time constants $\tau_i$ and $\tau_{rad}$, respectively. The resonator is coupled to free space via a single port, while the Bragg mirror is assumed to be a perfect reflector, preventing any transmission toward the substrate. This configuration has already been modeled using TCMT in previous works \cite{piper2014total,berguiga2021ultimate,Qu2015,Park2016}.

The relative reflection coefficient $r_{rel}(\omega)$, defined as the ratio of the reflection coefficient of the structure to that of the Bragg mirror, can be expressed as follows (see Supplementary Information for details):
\begin{equation}
r_{rel}(\omega)=\frac{\frac{1}{\tau_{rad}}-\frac{1}{\tau_i}-j\cdot \left(\omega-\omega_{res}\right)}{\frac{1}{\tau_{rad}}+\frac{1}{\tau_i}+j\cdot \left(\omega-\omega_{res}\right)}
\label{Equ_TMC_r_Tamm}
\end{equation}

From this equation, the reflection phase {$\phi_{rel}$} can be derived near resonance, assuming operation close to critical coupling ($\left|r_{min}\right|\ll1$). As shown in the Supplementary Information, we obtain:
\begin{equation}
\phi_{rel} \sim -2\cdot\frac{\omega-\omega_{res}}{\omega_{res}}\cdot\frac{Q}{r_{min}}+C
\label{Equ_TMC_phase1}
\end{equation}
with:
\begin{equation}
Q=\frac{\omega_{res}}{2}\cdot\frac{1}{\frac{1}{\tau_{rad}}+\frac{1}{\tau_i}} \text{\hspace{3ex} and \hspace{3ex}} r_{min}=\frac{\tau_{i}-\tau_{rad}}{\tau_{i}+\tau_{rad}}
\end{equation}

Here, $Q$ is the quality factor, and $r_{min}$ is the relative reflection amplitude at resonance. Equation \ref{Equ_TMC_phase1} shows that when $r_{min}=0$ (i.e., when $\tau_{rad}=\tau_{i}$), the relative phase shift becomes undefined. In other words, critical coupling is associated with a phase singularity. Moreover, the sign of the phase shift changes as the system crosses the critical point, since $r_{min}$ can be either positive or negative depending on the balance between radiative and internal decay rates. The constant C depends on the sign of $\tau_{rad}^{-1} - \tau_i^{-1}$: either the resonance corresponds to the under-coupling regime ($C$=0 when negative), either it corresponds to the over-coupling regime ($C$=$\pi$  when positive).

To obtain a sensing response, the phase shift $\phi_{rel}$ must be expressed as a function of the physical parameter of interest. For convenience, the following derivation is given for a thermal sensor with a temperature variation $dT$. However, the same reasoning applies to variations of analyte concentration $dC$ or refractive index $dn$. In a chemical sensor, the presence of an analyte induces a refractive index change only near the top surface, whereas in a thermal sensor, all refractive indices—including that of the sensor itself—may be affected. The phase sensitivity $S_{\phi}$ is then written as (see Supplementary Information for the full derivation):
\begin{equation}
S_{\phi}=\frac{d\phi_{rel}}{dT}=\frac{2}{\lambda_{res}}\cdot\frac{Q}{r_{min}}\cdot S_\lambda
\label{SensibilitevsLambda_2_Tamm2}
\end{equation}
where $\lambda_{res}$ is the resonance wavelength, and $S_\lambda=d\lambda_{res}/dT$ is the wavelength sensitivity. Equation \ref{SensibilitevsLambda_2_Tamm2} establishes a clear relationship between $S_{\phi}$, the wavelength sensitivity $S_{\lambda}$, the quality factor $Q$, and the minimum reflectivity $r_{min}$ for the considered system. In this work, the relationship previously defined for surface plasmon resonance in \cite{berguiga2021ultimate} is generalized to all systems where a single-mode resonator is coupled through a single port.

From Equation \ref{SensibilitevsLambda_2_Tamm2}, it is clear that $S_{\phi}$ scales linearly with $S_{\lambda}$: a good phase sensor must first be a good intensity sensor. Consequently, a high-performance phase sensor for chemical applications requires a maximum overlap between the electromagnetic field and the analyte, as is the case for any photonic sensor. In addition, two further guidelines can be emphasized for designing highly phase-sensitive sensors:
\begin{itemize}
\item Increasing the quality factor can drastically enhance phase sensitivity.
\item Lowering the reflectivity amplitude significantly increases phase sensitivity. Indeed, for a system close to critical coupling, the reflectivity amplitude can be reduced by several orders of magnitude, leading to extremely high phase sensitivity.
\end{itemize}

Our model demonstrates that the highest phase sensitivity is obtained when the reflectivity $|r_{min}|$ reaches its minimum.

\subsection{Tuning the reflectivity of nanopatterned optical Tamm structures} \label{tamm_expl}

The optical Tamm structure consists of a Distributed Bragg Reflector (DBR) (4 pairs of $Si/SiO_2$ layers with a $\lambda/4n$ thickness, designed to operate at $\lambda$=\SI{1.5}{\micro\meter}) covered by a thin gold layer (\SI{50}{\nano\meter}). This structure supports an optical Tamm mode at $\lambda$=\SI{1.68}{\micro\meter} in TE polarization, which can be excited at normal incidence. Such a mode generally exhibits a higher quality factor than a surface plasmon resonance due to its hybrid dielectric–metallic nature.

\begin{figure}[htb]
    \centering
   \includegraphics[width=1\linewidth]{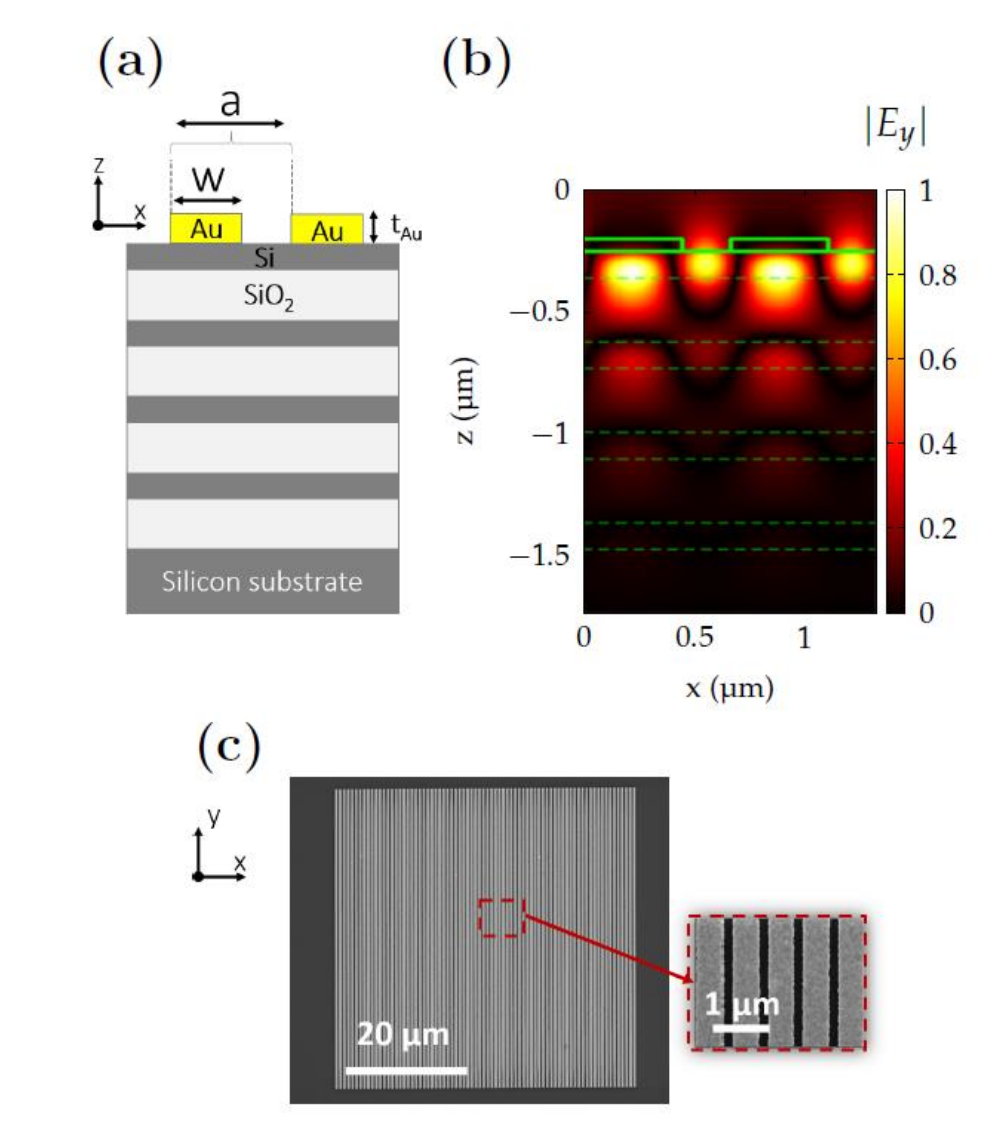}
    \caption{Simulated and fabricated Tamm structures consisting of a Bragg mirror (4 pairs $Si/SiO_2$) and a 1D gold grating. (a) Schematic of the optical Tamm structures with period $a$ and thickness $t_{Au}$. $w$ is the width of the gold rods. (b) Electric field magnitude $E_y$ with $a$~=~\SI{660}{\nano\meter}, $t_{Au}$~=~\SI{50}{\nano\meter}, and $w$~=~\SI{560}{\nano\meter}: green borders are the edges of the gold pattern and dashed lines locate the $Si/SiO_2$ interfaces. (c) Scanning Electron Microscope image of a fabricated 1D gold pattern.}
     \label{fig:fig1}
\end{figure}

Reaching the critical coupling condition in such structures requires design and, above all, fabrication at the nanometer scale \cite{auguie2015critical,zaky2020refractive,tsurimaki2018topological}. In particular, depositing the gold layer with sub-nanometric accuracy is necessary to approach the critical point, which is unachievable with conventional deposition techniques such as evaporation or sputtering. To overcome this limitation, the metallic layer is periodically patterned with a period $a$ and a metal rod width $w$ (figure \ref{fig:fig1} (a) et(b)). Each 1D pattern is a \SI{50}{}$\times$\SI{50}{\micro\meter} square with the same period ($a$=\SI{660}{\nano\meter}), but with varying $w$, which allows tuning of the resonance wavelength via the filling factor, defined as the ratio of the gold rod width $w$ to the period $a$ of the grating (FF~=~$w/a$) \cite{ferrier2019tamm}. Adjusting the filling factor tunes the balance between radiative and absorption losses \cite{auguie2015critical}. As discussed in the previous section, the minimum reflectivity (i.e., critical coupling) is determined by the balance of loss rates, so fine control of the filling factor allows resonances closer or farther from the critical point to be obtained.

\begin{figure*}[!ht]
\centering
   \includegraphics[width=1\linewidth]{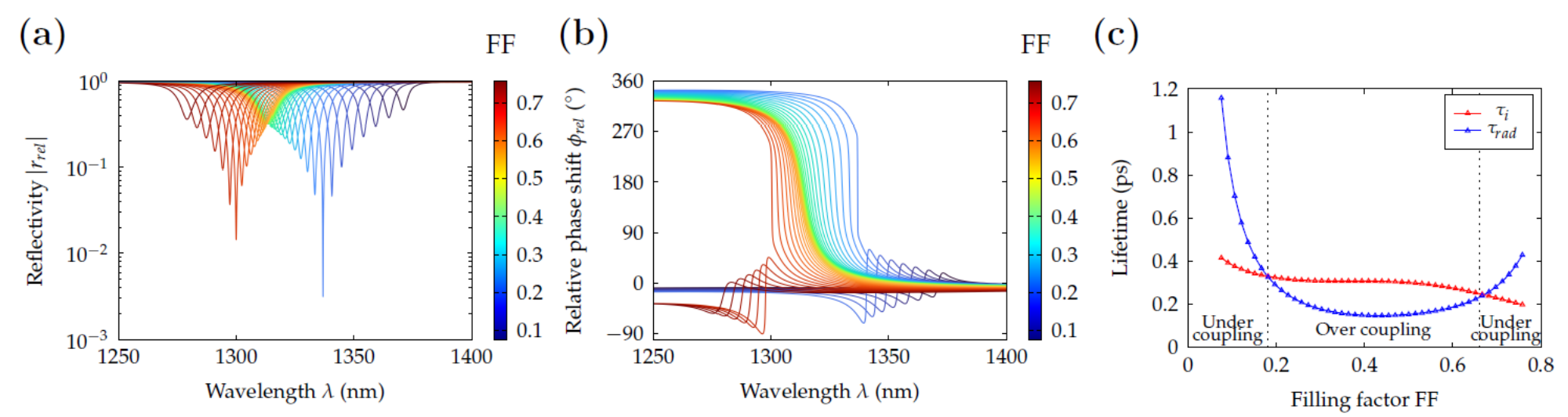}
    \caption{FDTD simulations of optical Tamm structures. All numerical simulations are performed with $a$=\SI{660}{\nano\meter} and $t_{Au}$=\SI{50}{\nano\meter} for s-polarized (TE) incident light. (a) Reflectivity amplitude. (b) Relative phase shift. (c) Values of $\tau_{rad}$ and $\tau_i$ calculated by fitting panels (a) and (b) using the complex reflection coefficient obtained from TCMT.}
    \label{fig:fig2}
\end{figure*}

To numerically validate this statement, we performed numerical simulations (FDTD) to retrieve the reflectivity minimum (amplitude) of the optical Tamm resonances as a function of the filling factor FF. Reflectivity spectra in TE polarization as a function of FF are shown in figure \ref{fig:fig2}(a), clearly demonstrating both resonance wavelength tuning and critical coupling tuning thanks to metal nanopatterning. Reflectivity values as low as $10^{-3}$ can be reached, and were achieved for two FF values, FF~=0.18 and FF=0.66. In figure \ref{fig:fig2}(b), the corresponding phase as a function of wavelength for each FF is plotted, showing the reflected phase associated with the Tamm resonance. Both critical coupling points exhibit abrupt phase jumps, corresponding to the transition between different operating regimes of the system: the under-coupled regime is characterized by a $\pi$ phase jump, whereas the over-coupled regime is characterized by a $2\pi$ phase jump. These different regimes and critical coupling points are dictated by the ratio between radiative and internal (absorption) losses. Indeed, the radiative losses are strongly modified by the patterning of the metal, whereas the absorption losses remain almost unchanged when varying the FF \cite{auguie2015critical}. This is clearly visible in figure \ref{fig:fig2}(c), where the evolution of radiative and internal losses is plotted as a function of FF (extraction of these values from FDTD numerical simulations is detailed in the Supplementary Information). Both critical points, where $\tau_{rad}$=$\tau_{i}$ and $R$$\rightarrow$$0$, are obtained, along with the different operating regimes of the structures: over-coupled ($\tau_{rad}$<$\tau_{i}$) and under-coupled ($\tau_{rad}$>~$\tau_{i}$). This is in clear agreement with a recent theoretical study predicting the number of spectral phase singularities in the case of a mirror-backed single-port system \cite{liu2023}.

Lastly, figure \ref{fig:fig2} illustrates that the slope of the phase shift $\phi_{rel}$ changes when transitioning from an over-coupling to an under-coupling regime, which theoretically changes the sign of $r_{min}$ and $S_{\phi}$.


\section{High phase sensitivity measurements}
\subsection{Fabrication of the samples} 
In order to experimentally demonstrate a highly sensitive phase sensor, we fabricated different patterned structures and characterized them thoroughly using intensity and phase measurements.
First, we deposited each layer of the Bragg mirror by Plasma Enhanced Chemical Vapor Deposition (PECVD) on a silicon substrate. Each layer was characterized by ellipsometry to control its refractive index and, in particular, its thickness, set to be $\lambda /4n$. The gold patterns on top of the Bragg mirror were defined by electron beam lithography, followed by a \SI{50}{\nano\meter} vapor deposition of gold and a lift-off process \cite{ferrier2019tamm}. A scanning electron microscopy image of one 1D grating is shown in figure \ref{fig:fig1}(c). The period of each 1D structure (\SI{50}{}$\times$\SI{50}{\micro\meter}) is \SI{660}{\nano\meter}. The FF is finely varied to obtain resonances as close as possible to the critical points. The resulting increase in the width of the gold rods from one structure to another is \SI{14}{\nano\meter}.

\subsection{Tuning of the critical coupling} \label{extraction}
\begin{figure}[!ht]
\centering
   \includegraphics[width=0.9\linewidth]{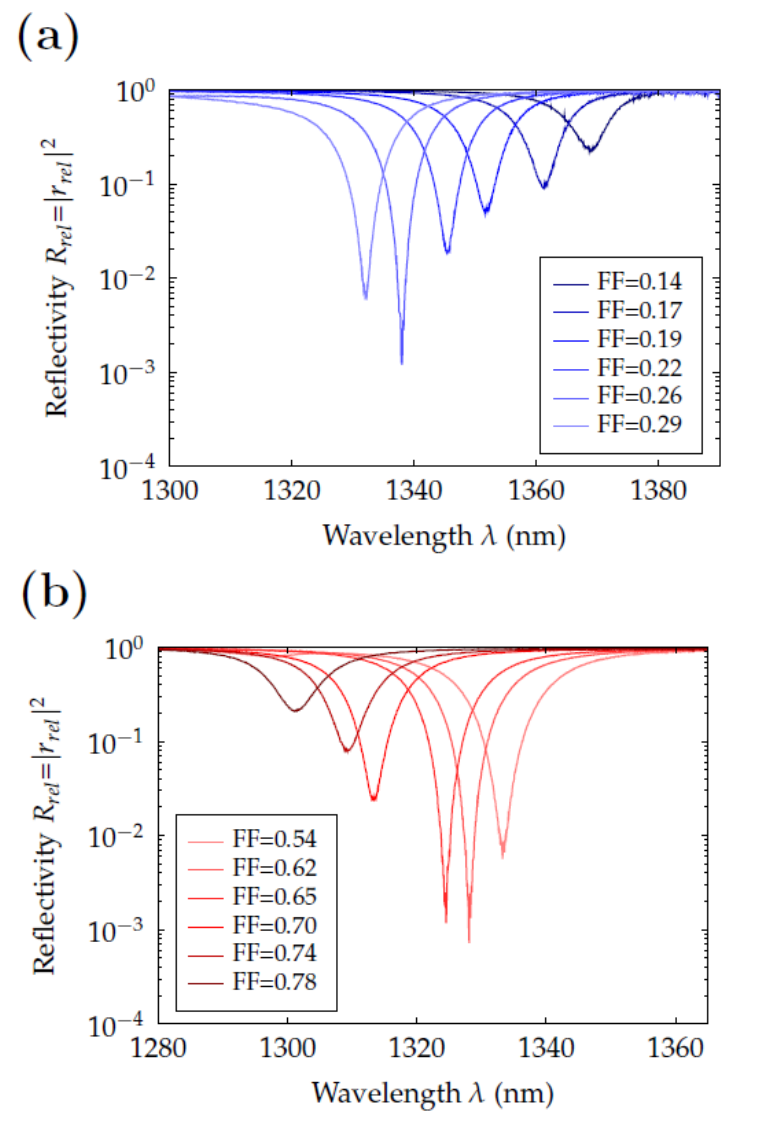}
     \caption{\label{fig:Rtot et Rmin(W)}Experimental reflectivity spectra as a function of the measured FF for $a$~=~\SI{660}{\nano\meter}. (a) FF around the first critical point and (b) around the second critical point.}
    \label{fig:fig3}
\end{figure}

Micro-reflectivity measurements were performed to demonstrate the experimental fine tuning of the critical coupling condition enabled by the 1D gold patterning. Spectra associated with different filling factors are presented in figure \ref{fig:fig3}(a) for low filling factors (0.11~<FF<~0.24) and in \ref{fig:fig3}(b) for high filling factors (0.46~<FF<~0.62). These experimental results show, first, that in good agreement with the FDTD simulation results presented in figure \ref{fig:fig2}, both the spectral position and the minimum reflectivity at resonance can be adjusted thanks to fine control of the filling factors. The small FF step between each structure allows us to decrease the minimum reflectivity by at least two orders of magnitude. These experimental data also allow us to extract the quality factors (around 100) and the experimental wavelength sensitivities $S_\lambda$ around the critical points when the sample is heated (see details in the Supplementary Information). With this set-up, we measure $S_\lambda$=\SI{0.117}{\nano\meter/\celsius}. Despite the low value (compared with the performance of other photonic crystal sensors \cite{zhou2020high,lu2019portable}), our purpose is to show that highly sensitive devices can be obtained by taking advantage of sharp phase jumps at critical coupling points. Measuring sharp phase variations and the resulting high phase sensitivities requires using an interferometric optical set-up designed for these measurements.

\subsection{Phase measurement optical set-up}
\begin{figure}[ht!]
\centering
   \includegraphics[width=1\linewidth]{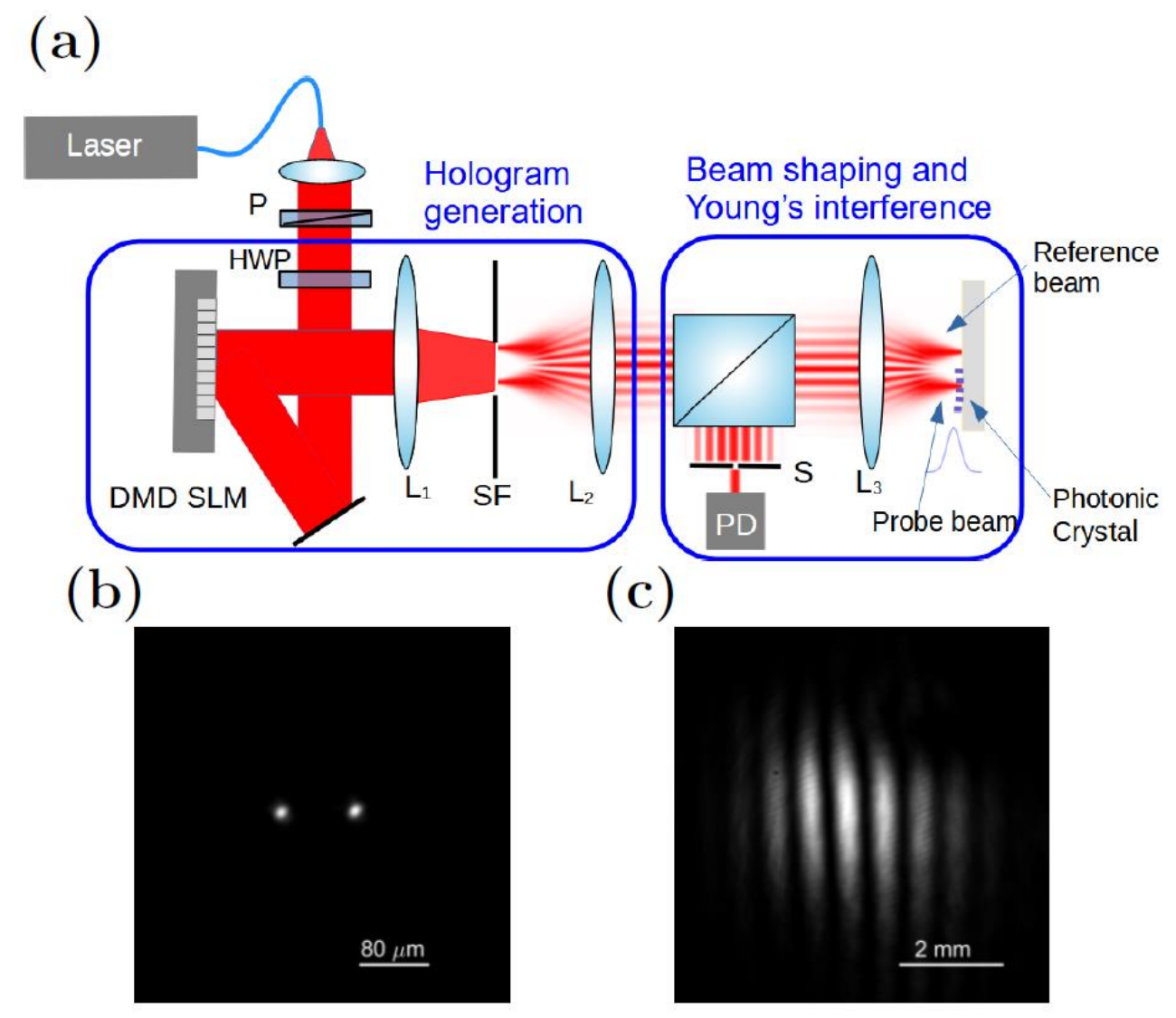}
     \caption{Optical setup for the phase sensitivity measurements. (a) Schematic of the setup. The light polarization is controlled using a polarizer ($P$) and a half-wave plate ($HWP$). Hologram generation and beam shaping are achieved with three lenses ($L_1$, $L_2$, $L_3$), a spatial filter (SF), and a spatial light modulator (SLM). The signal is detected after passing through a slit ($S$) of \SI{100}{\micro\meter} width in front of a photodetector (PD). (b) Camera image of the two Gaussian beams at the sample plane. (c) Young's interference pattern before the slit $S$.}
    \label{fig:fig4}
\end{figure}

An original common-path heterodyne digital interferometer, capable of measuring the phase with very high accuracy even for very low reflected light intensities, was used. This setup is described in detail in \cite{Girerd2024}. The general principle of the measurement is similar to a Young’s slit interference experiment: two beams are generated, one serving as the “reference” beam incident on the DBR, and the other as the “probe,” incident on the patterned structure. The interference pattern, resulting from the recombination of both beams, is recorded as a function of time. In this way, the phase can be measured with high accuracy, and sensors with very high sensitivity can be designed, as we will demonstrate later in the case of a temperature sensor. The setup integrates a holographic component that enables adequate shaping of the beams, which are adjusted to match the size of the Tamm structure. Figure \ref{fig:fig4}(a) shows a diagram of the optical setup.

The light from a tunable laser ($\lambda$=\SI{1260}{}–\SI{1360}{\nano\meter}), linearly polarized and controlled by a polarizer ($P$) and a half-wave plate ($HWP$) as detailed in figure \ref{fig:fig4}(a), passes through a system composed of a Digital Micromirror Device Spatial Light Modulator (DMD-SLM), three lenses ($L_1$, $L_2$, $L_3$), and a spatial filter (SF). The SLM allows manipulation of the incident wavefront by holography. Two Gaussian beams are shaped with the required characteristics (waist, separation distance, phase, and polarization). To achieve this, a hologram corresponding to the Fourier transform of the two Gaussian beams is first generated by the DMD-SLM; this hologram corresponds to a Young’s interference fringe pattern. Through the SLM, lenses $L_1$ and $L_2$, and the spatial filter (SF), the binary hologram is transformed into an analog one. Lens $L_3$ then generates the two beams that reach the sample. Both beams have a waist of \SI{8}{\micro\meter} and are separated center-to-center by \SI{75}{\micro\meter}. The probe beam excites the photonic crystal, while the second beam, the reference, is positioned nearby and illuminates the Bragg mirror. The reflected light from the sample is recombined by $L_3$, and the resulting interference pattern—after selecting a small portion of the central Young’s interference fringe with a slit $S$ of \SI{100}{\micro\meter} width—is measured with a photodetector (PD). To reduce phase noise and improve phase accuracy, one of the beams is phase-modulated in time at \SI{80}{\hertz} by the SLM \cite{Girerd2024}, resulting in a sinusoidal interferometric signal. Using Fourier analysis, the phase of the reflected light from the sensor is recovered by demodulation of this sinusoidal signal.

Our interferometric setup allows phase measurements for resonances with very low reflectivity amplitudes, down to $|r_{min}|$=\SI{e-3}{}. The slit in front of the photodetector selects only a portion of the emission angle of the reflected light, corresponding to a numerical aperture of \SI{0.014}{\degree}. This low value is essential to avoid flattening of the minimum reflectivity, which would otherwise increase the measured reflectivity and thereby reduce the performance of the interferometric setup (by yielding lower $S_{\phi}$ values than expected). Finally, a hot plate is used to control the temperature of the samples.

\subsection{Experimental critical coupling exploration}
\subsubsection{In wavelength}
\begin{figure*}[!ht]
\centering
   \includegraphics[width=1\linewidth]{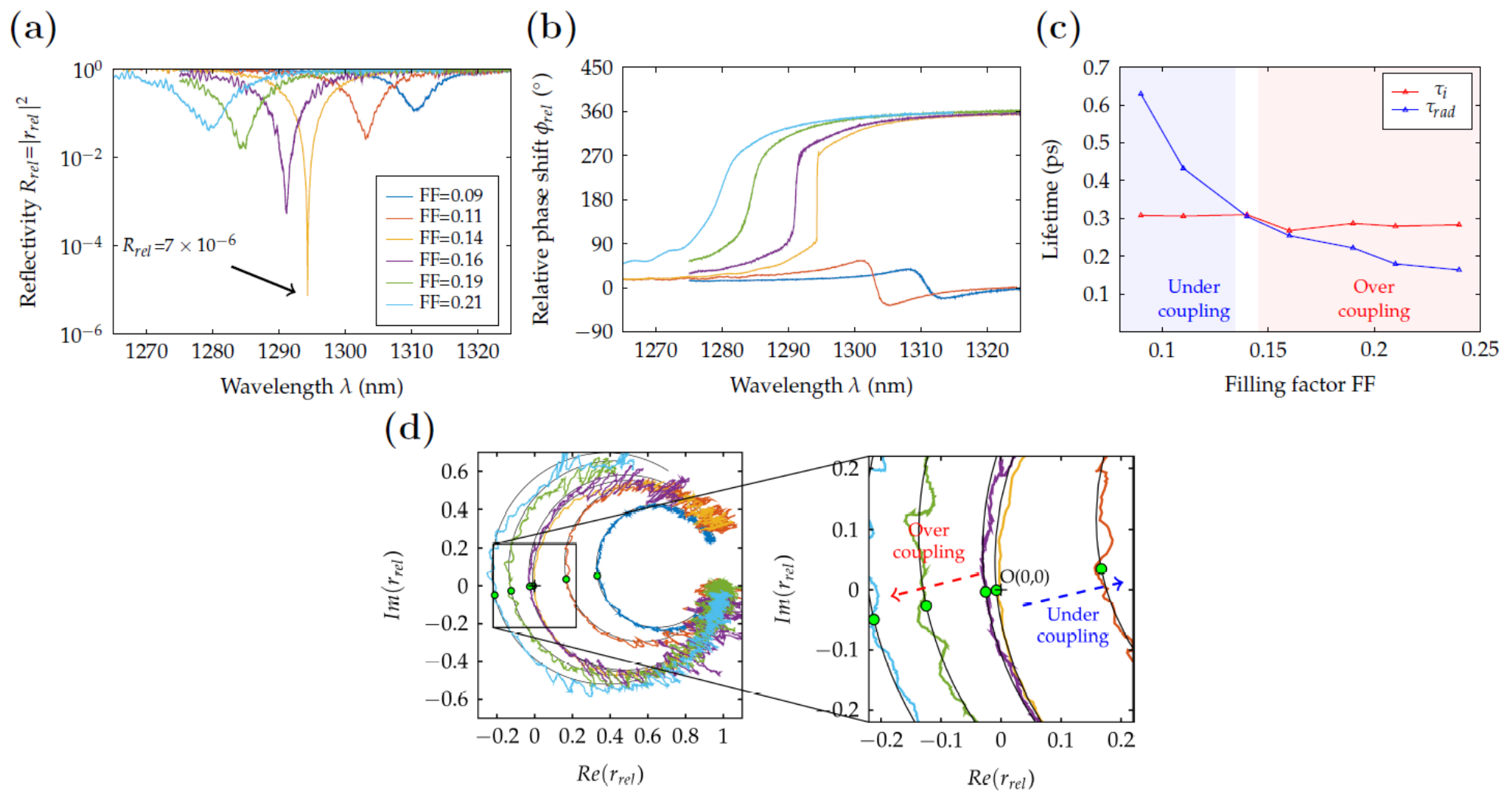}
    \caption{Experimental complex reflectivity spectra in intensity (a) and phase (b) for Tamm plasmon photonic crystals with different FF, with $a\approx$ \SI{600}{\nano\meter}. (c) Time constants (absorption and radiative losses) from the TMCT model are extracted by fitting the real and imaginary parts of the complex reflectivity versus the filling factor FF. (d) Representation of each spectrum in the complex plane. The resonance position (minimum of reflectivity) is marked with a green dot, whose distance from the origin O(0,0) is minimal. Each fit is represented by a dark thin circle.}
    \label{fig:fig5}
\end{figure*}

Critical coupling and its associated phase singularity are experimentally explored by measuring the complex reflectivity curves of several Tamm plasmon photonic crystals with filling factors ranging from 0.09 to 0.24. Figures \ref{fig:fig5}(a) and (b) show the intensity and phase reflectivity spectra for $FF$ values of 0.09, 0.11, 0.14, 0.16, 0.19, and 0.21. The reflectivity minimum, $R_{rel}$ = $|r_{rel}|^2$, reaches values as low as \SI{7e-6}{} for $FF$=0.14. Figure \ref{fig:fig5}(b) unambiguously shows the experimental transition from under- to over-coupling, as well as the abrupt phase change close to the critical coupling point. The time constants $\tau_{rad}$ and $\tau_{i}$ for each photonic crystal are extracted by fitting the real and imaginary parts of the reflectivity spectra $r_{rel}$ using the TMCT equation \ref{Equ_TMC_r_Tamm} (calculation details for each $FF$ are provided in the Supplementary Material). The evolution of both time constants with $FF$ is shown in figure \ref{fig:fig5}(c): as $FF$ increases, the experimental radiative time constant $\tau_{rad}$ decreases from about \SI{0.60}{\pico\second} to \SI{0.16}{\pico\second}, whereas the intrinsic time constant $\tau_{i}$, related to absorption losses, remains nearly constant around \SI{0.29}{\pico\second}–\SI{0.30}{\pico\second}. Both time constants intersect for $FF$ values around 0.15, corresponding to extremely low reflected optical intensity and sharp phase variations. Under-coupling ($\tau_{rad}$>$\tau_{i}$) occurs when $FF$ is below 0.14, whereas over-coupling ($\tau_{rad}$<$\tau_{i}$) occurs above 0.16. These experimental measurements confirm the numerical simulations presented in figure \ref{fig:fig2}(c), the theoretical work on thin-film Tamm plasmons \cite{auguie2015critical}, and the theoretical and experimental studies in the mid-infrared domain by Zhou’s group \cite{Qu2015}.

Another representation of the resonance in the complex plane provides a direct overview of the passage through critical coupling (figures \ref{fig:fig5}(d)). Each measured spectrum is represented by a circle, where the distance of each point from the origin O(0,0) corresponds to the modulus of the reflectivity $|r_{rel}|$ for a given wavelength. The minimum reflectivity $|r_{min}|$ is given by the point on the circle closest to the origin, and critical coupling is approached when this closest point converges toward the origin. The under-coupling regime occurs when the origin lies outside the circle, whereas over-coupling occurs when the origin lies inside the circle (see the zoom in figure \ref{fig:fig5}(d)).

\subsubsection{In wavevector space (angular distribution)}

\begin{figure*}[htb]
    \centering
    \includegraphics[width=0.8\linewidth]{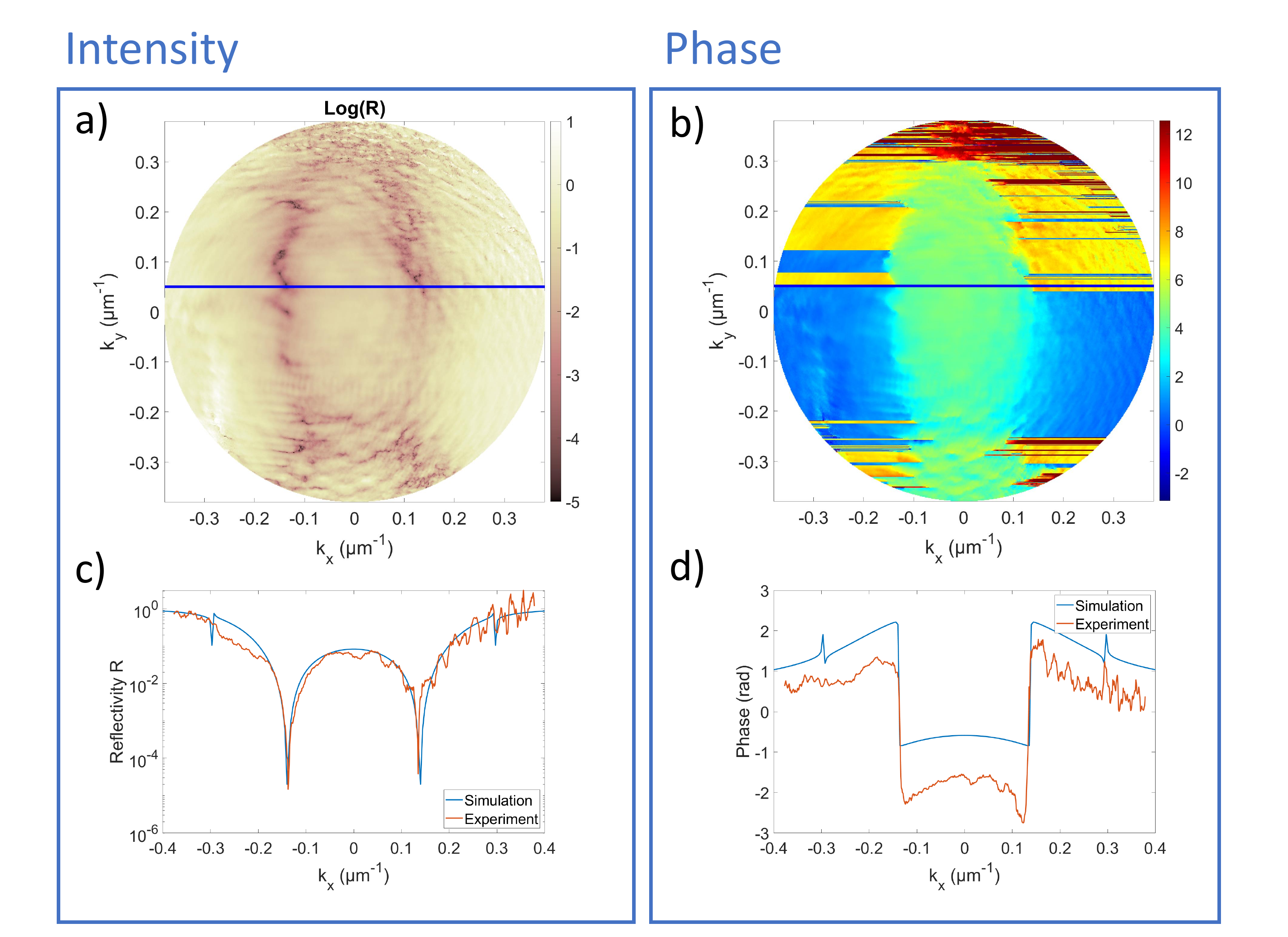}
    \caption{Experimental complex reflectivity distribution at a wavelength of 1298.38 nm for a Tamm plasmon structure with a=\SI{600}{\nano\meter} and $FF$=0.14. (a) and (b) show the intensity $Log(R_{rel})$ and phase $\phi_{rel}$, respectively. The orange curves in (c) and (d) correspond to sections along the blue and black lines in (a) and (b). The blue curves in (c) and (d) are simulations for a Tamm plasmon structure with $a=600$ nm, $FF=0.138$, and a wavelength of 1294.25 nm at $k_y=0$.}
    \label{fig:fig6}
\end{figure*}

If the wavelength is fixed at resonance for a Tamm plasmon structure (with $FF$=0.14 and a=\SI{600}{\nano\meter}) and the interference pattern is recorded by an IR camera instead of the slit and photodiode—while using the same demodulation process—two images are obtained, showing the intensity and phase reflectivity ($R_{rel}$ and $\phi_{rel}$) as functions of the wavevectors $k_x$ and $k_y$, as shown in figures \ref{fig:fig6}(a) and b). In figure \ref{fig:fig6}(a), two nearly vertical symmetrical lines are observed around the wavevector $k_x$=$\pm$\SI{0.135}{\per\micro\meter} (corresponding to an angle $\theta_x=\pm 1.65^{\circ}$) with reflectivity lower than $10^{-3}$. In the phase map, this corresponds to a sharp jump (figure \ref{fig:fig6}(b)).
Figures \ref{fig:fig6}(c) and (d) show sections of images a) and b) along the blue and black lines, respectively, which are well fitted by simulations with $FF=0.138$ and a wavelength of \SI{1294.25}{\nano\meter}. These two lines therefore mark the wavevector positions of the critical coupling and its phase singularity.
The slight curvature of the symmetrical vertical lines associated with the phase singularity in figures \ref{fig:fig6}(a) and b) is not an optical aberration but rather a result of the optical response of the photonic crystal, as confirmed by numerical simulations in the Supplementary Data.

These results show that the phase singularity does not occur at normal incidence (the $\Gamma$ point) but at a finite angle. In fact, it is experimentally very difficult to obtain the precise $FF$ that produces the phase singularity at normal incidence. Numerical simulations show that as $FF$ is finely tuned, the vertical lines of the phase singularity in figures \ref{fig:fig6}(a) and (b) move symmetrically toward the central position $k_x=0$ and eventually merge (see Supplementary Data). This means that, for sensor measurements and for achieving the most sensitive photonic crystal response, the slit and photodiode must be carefully positioned at the correct wavevector.

\subsection{Phase sensitivity measurements}
\begin{figure*}[ht!]
\centering
    \includegraphics[width=1\linewidth]{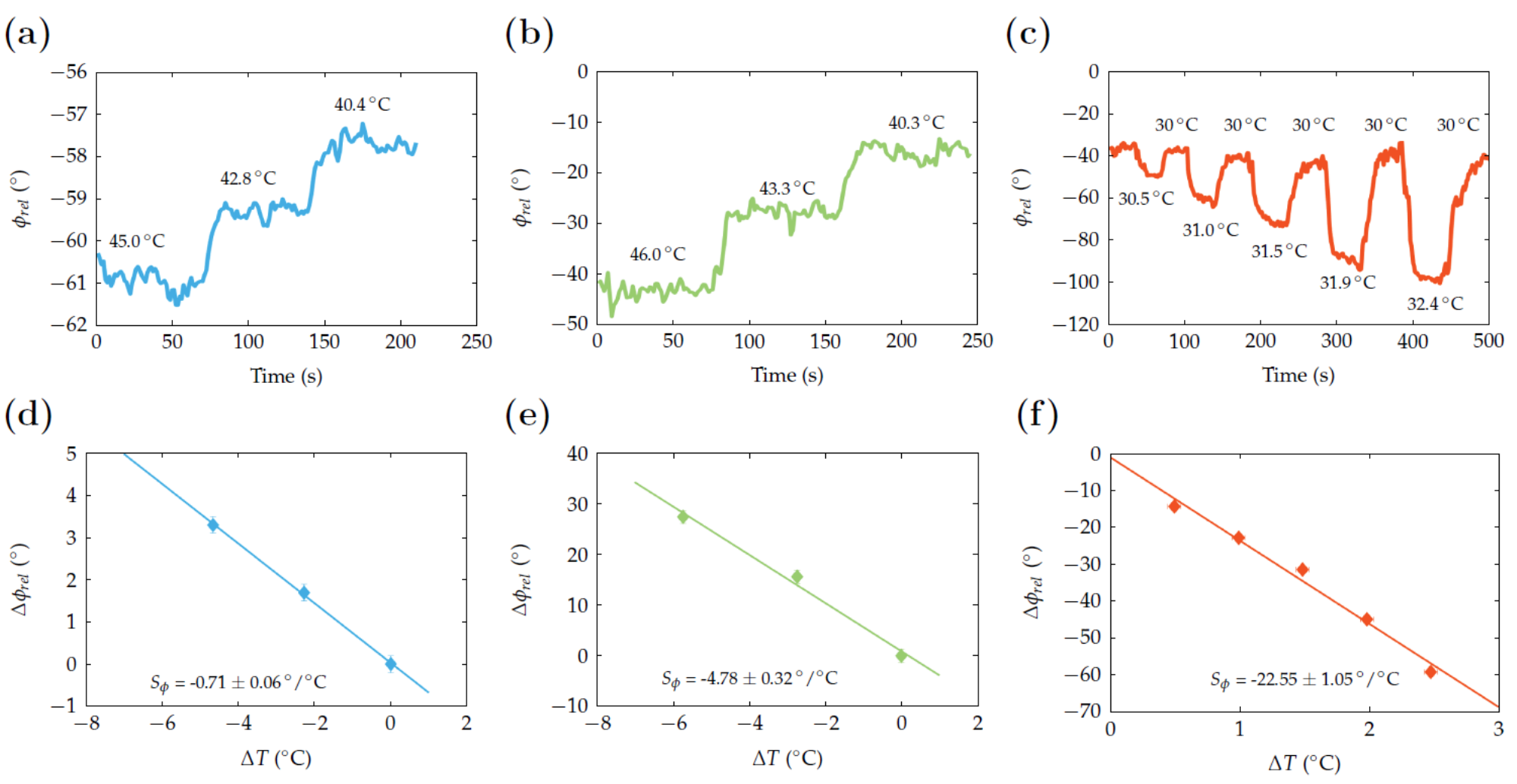}
\caption{Phase of the reflected light from different Tamm structures as a function of time for various heating or cooling cycles. Each structure has a period of $a$=\SI{660}{\nano\meter}. Measurements were obtained on a structure with FF~=0.78 ($R_{min}$=0.24, i.e., $|r_{min}|$=0.49) for (a), with FF=0.72 ($R_{min}$=0.012, i.e., $|r_{min}|$=0.11) for (b), and with FF=0.69 ($R_{min}$=0.002, i.e., $|r_{min}|$=0.045) for (c). Panels (d)–(f) show the phase variation $\Delta\phi_{rel}$ versus temperature variation $\Delta T$, respectively extracted from (a), (b), and (c), with the reference phase and temperature taken at t=~\SI{0}{\second}. The phase sensitivity $S_{\phi}$ corresponds to the slope of the line fitting the data}
    \label{fig:fig7}
\end{figure*}

The sample is placed in the experimental setup and illuminated at a fixed wavelength corresponding to the reflectivity minimum of the photonic crystal under study. As previously explained, the phase shift $\phi_{rel}$ of the reflected light is recorded as a function of time. Figures \ref{fig:fig7}(a), \ref{fig:fig7}(b), and \ref{fig:fig7}(c) show examples of phase jumps measured with the interferometric setup for Tamm structures with different $FF$ values (i.e., different $r_{min}$). The sample is either cooled from \SI{46}{\celsius} to \SI{40}{\celsius} (figures \ref{fig:fig7}(a) and \ref{fig:fig7}(b)), or periodically heated and cooled with an incremental step of \SI{0.5}{\celsius} between successive heating cycles (figure \ref{fig:fig7}(c)).

The average phase jump $\Delta\phi_{rel}$ is plotted as a function of the temperature step $\Delta T=T_{final}-T_{t=\SI{0}{\second}}$ in figures \ref{fig:fig7}(d)-(e), showing that the phase shift evolves linearly with temperature. The phase sensitivity of each structure is extracted from the slope of each curve.

\begin{figure}[ht!]
    \centering
	\includegraphics[width=1\linewidth]{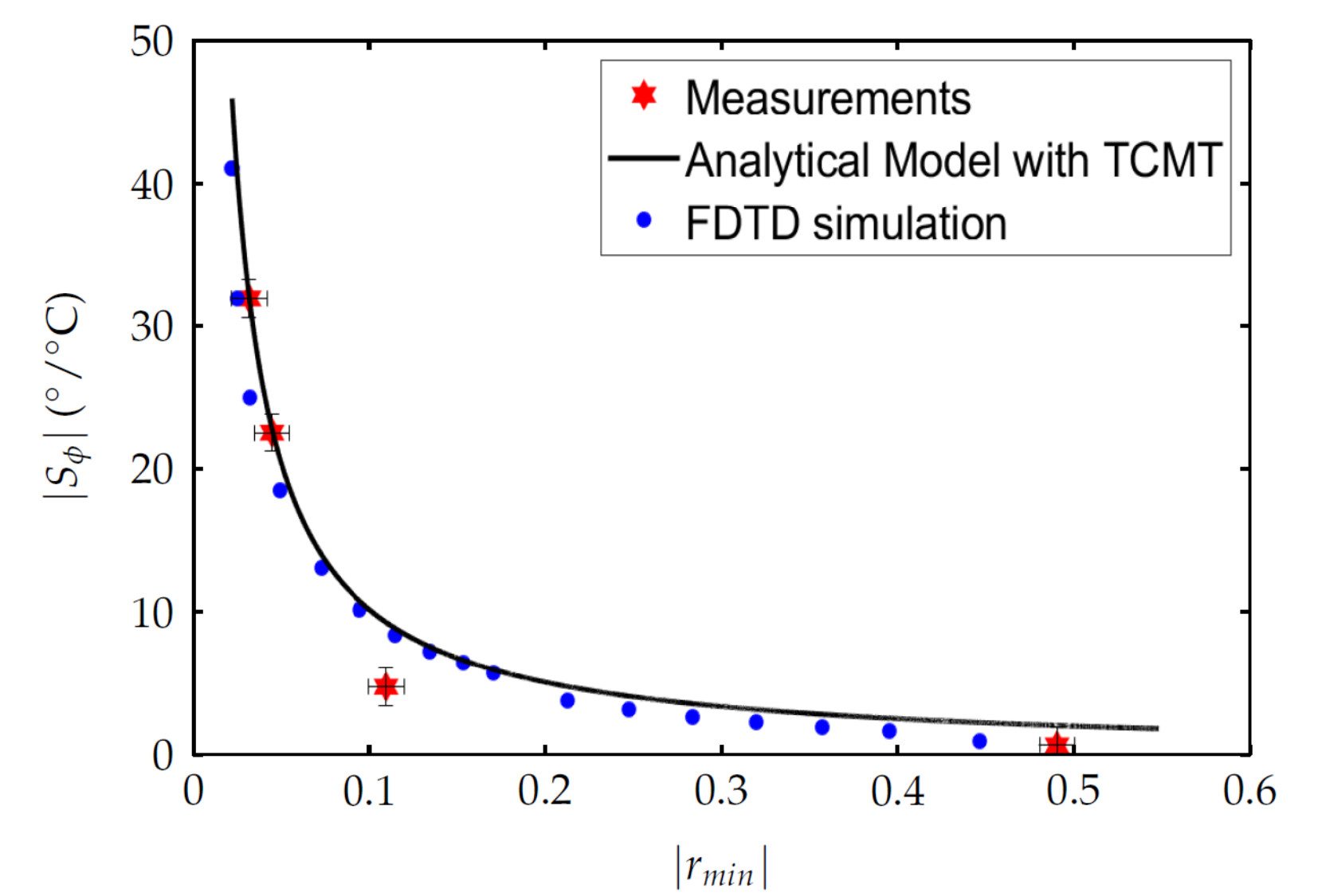}
    \caption{Phase sensitivity as a function of the minimum reflectivity: comparison of experimental measurements, FDTD simulations, and the theoretical model from the equation \ref{SensibilitevsLambda_2_Tamm2}}.
    \label{fig:fig8}
\end{figure}

The measured phase sensitivities and their associated uncertainties, as a function of $|r_{min}|$, are shown as red stars in figure \ref{fig:fig8}. To demonstrate the good agreement between the phase sensitivity measurements, FDTD numerical simulations, and the analytical model developed in section \ref{tamm_expl}, the numerical and analytical results are also plotted in figure \ref{fig:fig8}. The experimental data for the quality factor $Q$ and the wavelength sensitivity $S_{\lambda}$, obtained from intensity measurements in section \ref{extraction}, are implemented in equation \ref{SensibilitevsLambda_2_Tamm2} to plot the phase sensitivity as a function of $|r_{min}|$. In other words, no fitting parameters were introduced in generating this curve.

As expected from equation (\ref{SensibilitevsLambda_2_Tamm2}), the decrease of the reflection coefficient $|r_{min}|$, due to the critical coupling phenomenon, results in a drastic increase of the phase sensitivity—by more than one order of magnitude—clearly highlighted in the experimental results. Figure \ref{fig:fig8} shows that the model developed from TCMT accurately describes the behavior of the phase sensitivity of optical modes near the critical coupling regime. It also demonstrates that we can tune the filling factor of the nanostructures to achieve sensors with very high sensitivity, and that the common-path interferometric setup based on digital holography is sufficiently precise to measure this sensitivity.

\subsection{Limit of detection}

\begin{figure}[htb]
    \centering
    \includegraphics[width=1\linewidth]{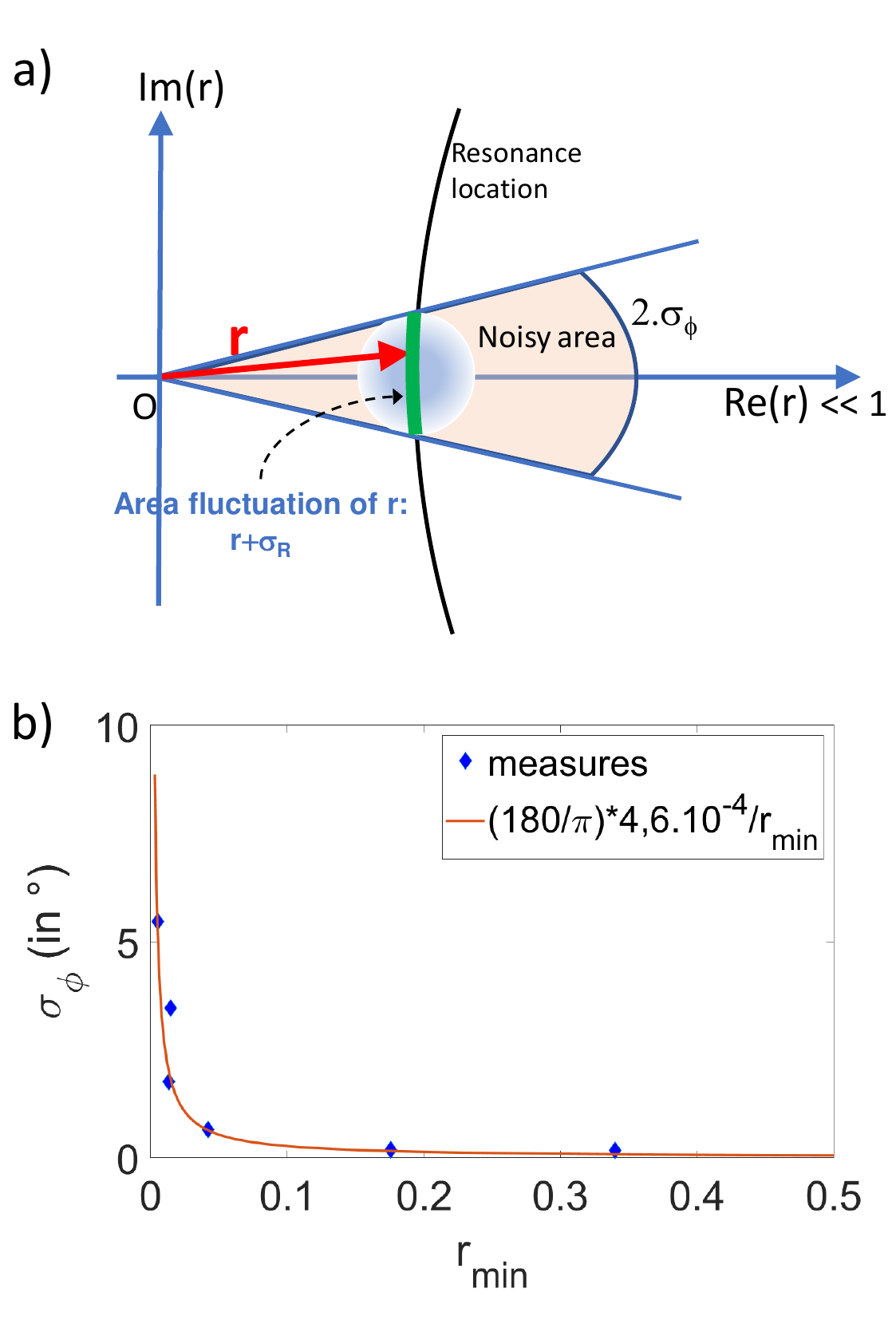}
    \caption{a) Geometric interpretation of the intensity and phase noise ($\sigma_R$ and $\sigma_\phi$). b) Experimental evolution of the phase noise as a function of the minimum reflectivity $r_{min}$. The theoretical curve, following the law $\sigma_R/r_{min}$, is plotted using the experimental value of $\sigma_R = 4.6 \times 10^{-4}$.}
    \label{fig:fig9}
\end{figure}

The limit of detection (LOD) is a key parameter in sensing applications, defining the ability of a sensor to achieve high resolution and very low detection thresholds. The LOD is determined by the sensitivity of the sensor and the measurement noise. In our phase interrogation method, the LOD is defined as:
\begin{equation}
LOD_{\phi}=3\cdot\frac{\sigma_{\phi}}{S_{\phi}}
\label{eq:LODdef}
\end{equation}
where $\sigma_{\phi}$ is the phase noise of the experimental set-up. To compare the phase noise with the more commonly used parameter, the intensity noise $\sigma_{R}$, we must establish a relationship between them. In spectral interrogation, Konteduca et al. \cite{Conteduca2022} derived a relation between their LOD and the intensity noise $\sigma_{R}$ using TCMT. In our case, to relate the phase noise to the intensity noise, we represent the real and imaginary parts of the reflectance in the complex plane, with $r$ as a vector originating from O(0,0). For random noise fluctuations, the tip of the vector $r$ evolves within a disc of radius $\sigma_R$, as shown by the blue disc in figure \ref{fig:fig9}(a). Since reflectivity is a resonant phenomenon, the vector $r$ is constrained to evolve on a circle corresponding to the resonance. Consequently, the vector $r$ fluctuates along a limited arc, as shown by the green curve in figure \ref{fig:fig9}(a). This fluctuation leads to an angular variation and therefore to a phase fluctuation $2\sigma_{\phi}$, governed by the size of this arc. As the critical coupling is approached, the vector $r$ becomes smaller, and the blue disc and green arc move closer to the origin O, leading to a larger cone angle. By construction, the relationship between phase noise and intensity noise is:
\begin{equation}
\sigma_{\phi}=\frac{\sigma_R}{r_{min}+\frac{\sigma_R^2}{2\cdot R_o}}\sim\frac{\sigma_R}{r_{min}}
\label{eq:PhaseNoiseRelation}
\end{equation}
where $R_o$ is the radius of the complex resonance circle, which is close to 0.5 near the critical coupling. The term $\frac{\sigma_R^2}{2.R_o}$, related to the circle curvature, is negligible compared to $r_{min}$ because $\sigma_R\ll1$.

We verified this relationship experimentally by measuring amplitude and phase stability for four different Tamm plasmon structures with different values of $r_{min}$ at resonance (see supplementary data). Figure \ref{fig:fig9}(b) shows the evolution of phase noise versus $r_{min}$, which can be fitted by relation (\ref{eq:PhaseNoiseRelation}) using $\sigma_R=4.6\times10^{-4}$. This intensity noise value is consistent with the measured standard deviation of the reflectivity (see supplementary data), which ranges between $2\times10^{-4}$ and $8\times10^{-4}$. It is also consistent with the background value of the power spectral density of the modulated interferometric signal \cite{Girerd2024}. No assumption was made regarding the origin of the intensity noise.

It follows that phase noise increases proportionally to the inverse of $r_{min}$. In other words, approaching critical coupling degrades the precision of phase measurements. From equations (\ref{SensibilitevsLambda_2_Tamm2}) and (\ref{eq:LODdef}), the LOD becomes:
\begin{equation}
LOD_{\phi}=\frac{3}{2}\cdot\frac{\lambda_{res}}{Q\cdot S_{\lambda}}\cdot\sigma_R
\label{eq:LODexpression}
\end{equation}

The LOD obtained from figure \ref{fig:fig7} for the most sensitive sensor is 0.078 $^{\circ}$C, which is very close to the value predicted by equation (\ref{eq:LODexpression}): 0.063 $^{\circ}$C. This expression shows that it is not beneficial to approach critical coupling to improve the detection limit, since the LOD is independent of the minimum reflectivity. As the critical coupling regime is approached, the enhancement in phase sensitivity is counterbalanced by an associated increase in phase noise.

\section{Discussion}
It is well established that approaching critical coupling and its associated phase singularity enhances phase sensitivity \cite{tsurimaki2018topological,vasic2014enhanced,nikitin1999surface,ermolaev2022topological,ma2023excitons,Wang2024}. In the works of Ermolaev \cite{ermolaev2022topological} and Ma \cite{ma2023excitons}, very high phase sensitivities were achieved by ellipsometry, ranging from 2 to $7 \times 10^4$ $^{\circ}/RIU$, thanks to the use of transition metal dichalcogenides, which enable the generation of topological phase singularities. The role of the quality factor, wavelength sensitivity, and minimum reflectivity $|r_{min}|$ has been identified in earlier studies \cite{tsurimaki2018topological,vasic2014enhanced,nikitin1999surface}, but without a complete analytical framework.
In this work, using TCMT modeling, we generalized the analytical relation between $S_{\phi}$ and $r_{min}$, previously derived for surface plasmon resonance in \cite{berguiga2021ultimate}. We highlight the role of all these parameters for such structures and experimentally validate this relation through careful design and fabrication, exploring phase sensitivity around critical points.

From equation (\ref{SensibilitevsLambda_2_Tamm2}), it follows that the only parameters linking phase to wavelength sensitivity are the resonance width $\Delta\lambda_{res}$ and the minimum reflectivity $r_{min}$, i.e. the two key parameters that describe the resonance itself. Indeed, this equation can also be written as:

\begin{equation}
\left|S_{\phi}\right|=\frac{2}{\Delta\lambda_{res}\cdot |r_{min}|}\cdot \left|S_\lambda\right|
\end{equation}

The importance of phase singularities in enhancing phase sensitivity becomes even clearer when equation (\ref{SensibilitevsLambda_2_Tamm2}) is rewritten in terms of the internal and radiative time constants of the sensor, $\tau_i$ and $\tau_{rad}$:

\begin{equation}
S_{\phi}=\frac{2\pi\cdot c}{\lambda_{res}^2}\cdot\frac{\tau_i\cdot \tau_{rad}}{\tau_i-\tau_{rad}}\cdot S_{\lambda}
\label{SensibilitevsLambda_2_Tamm_ecriture2}
\end{equation}

This equation shows that achieving very high phase sensitivities does not simply require low losses (high $Q$, i.e. both $\tau_{rad}$ and $\tau_i$ large), but more importantly, a small difference between these two loss channels. In other words, proximity to the critical coupling point, where $\tau_{rad}$ and $\tau_i$ match, leads to extremely high sensitivities and even divergence.

From a photonics perspective, the equations for phase sensitivity and LOD can be generalized to any single-mode resonator operating close to a critical point (surface plasmon resonance, Tamm plasmon modes, Bloch surface waves, etc.). Moreover, equation (\ref{SensibilitevsLambda_2_Tamm2}), which describes the phase sensitivity of a photonic sensor, can be extended to refractive index sensors such as gas or biomolecular detectors (see equation (16) in the Supplemental Material, derived for variations in the surrounding refractive index).

Although previous works \cite{tsurimaki2018topological,vasic2014enhanced,nikitin1999surface,ermolaev2022topological,ma2023excitons,Wang2024} reported enhanced phase sensitivity near phase singularities, to the best of our knowledge, no study has discussed the limit of detection (LOD), which determines the true efficiency of a photonic sensor. Here, we experimentally demonstrate for the first time that in phase interrogation (ellipsometry or interferometry), the LOD is not governed by critical coupling, as shown in equation (\ref{eq:LODexpression}). Approaching critical coupling and its associated phase singularity does not reduce the LOD, contrary to the common assumption for sensors exploiting phase singularities. The typically low LOD observed in phase interrogation methods is therefore not a direct consequence of phase singularity.

We show that the phase singularity itself has no effect on the LOD of phase sensors operating at critical coupling. However, in metasurfaces exhibiting vortex phase singularities with high topological charge \cite{liu2023,Zhen2014}, the phase sensitivity is expected to be enhanced by a factor equal to the topological charge $q$. The LOD would also scale with the same factor, assuming no additional noise arises from positioning the detection slit around the vortex. This hypothesis should be verified experimentally. If confirmed, it would open a promising route toward improving the LOD of photonic structures that combine phase singularities with high topological charge.

\section{Conclusion}
In conclusion, we have been able to explore the critical coupling point very closely thanks to precise control of the geometry of the Tamm plasmon nanostructures during fabrication. Using the interferometric measurement setup, we measured very low reflectivity and sharp phase jumps at resonance due to the phase singularity. This work provides a precise and comprehensive theoretical and experimental understanding of photonic sensors operating at critical coupling. We show that the phase sensitivity increases drastically as the critical coupling and its phase singularity are approached, and we determine all the key physical parameters quantitatively. We also demonstrate that approaching the phase singularity has no effect on the detection limit and therefore does not improve it. In fact, the phase singularity is not the origin of the low detection limits typically observed in phase interrogation methods. We propose that more sensitive sensors with low detection limits can be achieved by significantly increasing the topological charge of phase singularities. Consequently, phase singularities remain of great interest in cases where the singularities have a high topological charge. This represents a promising avenue for designing photonic sensors with high topological charges and experimentally verifying this hypothesis.

\section{Funding}
This work  was supported by Rhône Alpes Auvergne Region under the project INSIDEPACK+ (grant RA 0030195 beneath the "Fonds Européen de Développement Régional (FEDER)".)


\bibliography{biblio}

\begin{thebibliography}{52}%
\makeatletter
\providecommand \@ifxundefined [1]{%
 \@ifx{#1\undefined}
}%
\providecommand \@ifnum [1]{%
 \ifnum #1\expandafter \@firstoftwo
 \else \expandafter \@secondoftwo
 \fi
}%
\providecommand \@ifx [1]{%
 \ifx #1\expandafter \@firstoftwo
 \else \expandafter \@secondoftwo
 \fi
}%
\providecommand \natexlab [1]{#1}%
\providecommand \enquote  [1]{``#1''}%
\providecommand \bibnamefont  [1]{#1}%
\providecommand \bibfnamefont [1]{#1}%
\providecommand \citenamefont [1]{#1}%
\providecommand \href@noop [0]{\@secondoftwo}%
\providecommand \href [0]{\begingroup \@sanitize@url \@href}%
\providecommand \@href[1]{\@@startlink{#1}\@@href}%
\providecommand \@@href[1]{\endgroup#1\@@endlink}%
\providecommand \@sanitize@url [0]{\catcode `\\12\catcode `\$12\catcode `\&12\catcode `\#12\catcode `\^12\catcode `\_12\catcode `\%12\relax}%
\providecommand \@@startlink[1]{}%
\providecommand \@@endlink[0]{}%
\providecommand \url  [0]{\begingroup\@sanitize@url \@url }%
\providecommand \@url [1]{\endgroup\@href {#1}{\urlprefix }}%
\providecommand \urlprefix  [0]{URL }%
\providecommand \Eprint [0]{\href }%
\providecommand \doibase [0]{https://doi.org/}%
\providecommand \selectlanguage [0]{\@gobble}%
\providecommand \bibinfo  [0]{\@secondoftwo}%
\providecommand \bibfield  [0]{\@secondoftwo}%
\providecommand \translation [1]{[#1]}%
\providecommand \BibitemOpen [0]{}%
\providecommand \bibitemStop [0]{}%
\providecommand \bibitemNoStop [0]{.\EOS\space}%
\providecommand \EOS [0]{\spacefactor3000\relax}%
\providecommand \BibitemShut  [1]{\csname bibitem#1\endcsname}%
\let\auto@bib@innerbib\@empty
\bibitem [{\citenamefont {Blevins}\ \emph {et~al.}(2021)\citenamefont {Blevins}, \citenamefont {Fernandez-Galiana}, \citenamefont {Hooper},\ and\ \citenamefont {Boriskina}}]{blevins2021roadmap}%
  \BibitemOpen
  \bibfield  {author} {\bibinfo {author} {\bibfnamefont {M.~G.}\ \bibnamefont {Blevins}}, \bibinfo {author} {\bibfnamefont {A.}~\bibnamefont {Fernandez-Galiana}}, \bibinfo {author} {\bibfnamefont {M.~J.}\ \bibnamefont {Hooper}},\ and\ \bibinfo {author} {\bibfnamefont {S.~V.}\ \bibnamefont {Boriskina}},\ }\bibfield  {title} {\bibinfo {title} {Roadmap on universal photonic biosensors for real-time detection of emerging pathogens},\ }in\ \href@noop {} {\emph {\bibinfo {booktitle} {Photonics}}},\ Vol.~\bibinfo {volume} {8}\ (\bibinfo {organization} {MDPI},\ \bibinfo {year} {2021})\ p.\ \bibinfo {pages} {342}\BibitemShut {NoStop}%
\bibitem [{\citenamefont {Kaur}\ \emph {et~al.}(2022)\citenamefont {Kaur}, \citenamefont {Kumar},\ and\ \citenamefont {Kaushik}}]{kaur2022recent}%
  \BibitemOpen
  \bibfield  {author} {\bibinfo {author} {\bibfnamefont {B.}~\bibnamefont {Kaur}}, \bibinfo {author} {\bibfnamefont {S.}~\bibnamefont {Kumar}},\ and\ \bibinfo {author} {\bibfnamefont {B.~K.}\ \bibnamefont {Kaushik}},\ }\bibfield  {title} {\bibinfo {title} {Recent advancements in optical biosensors for cancer detection},\ }\href@noop {} {\bibfield  {journal} {\bibinfo  {journal} {Biosensors and Bioelectronics}\ }\textbf {\bibinfo {volume} {197}},\ \bibinfo {pages} {113805} (\bibinfo {year} {2022})}\BibitemShut {NoStop}%
\bibitem [{\citenamefont {Parvin}\ \emph {et~al.}(2021)\citenamefont {Parvin}, \citenamefont {Ahmed}, \citenamefont {Alatwi},\ and\ \citenamefont {Rashed}}]{parvin2021differential}%
  \BibitemOpen
  \bibfield  {author} {\bibinfo {author} {\bibfnamefont {T.}~\bibnamefont {Parvin}}, \bibinfo {author} {\bibfnamefont {K.}~\bibnamefont {Ahmed}}, \bibinfo {author} {\bibfnamefont {A.~M.}\ \bibnamefont {Alatwi}},\ and\ \bibinfo {author} {\bibfnamefont {A.~N.~Z.}\ \bibnamefont {Rashed}},\ }\bibfield  {title} {\bibinfo {title} {Differential optical absorption spectroscopy-based refractive index sensor for cancer cell detection},\ }\href@noop {} {\bibfield  {journal} {\bibinfo  {journal} {Optical Review}\ }\textbf {\bibinfo {volume} {28}},\ \bibinfo {pages} {134} (\bibinfo {year} {2021})}\BibitemShut {NoStop}%
\bibitem [{\citenamefont {Gallego}\ \emph {et~al.}(2008)\citenamefont {Gallego}, \citenamefont {Roca}, \citenamefont {Guardino},\ and\ \citenamefont {Rosell}}]{gallego2008indoor}%
  \BibitemOpen
  \bibfield  {author} {\bibinfo {author} {\bibfnamefont {E.}~\bibnamefont {Gallego}}, \bibinfo {author} {\bibfnamefont {F.}~\bibnamefont {Roca}}, \bibinfo {author} {\bibfnamefont {X.}~\bibnamefont {Guardino}},\ and\ \bibinfo {author} {\bibfnamefont {M.}~\bibnamefont {Rosell}},\ }\bibfield  {title} {\bibinfo {title} {{Indoor and outdoor BTX levels in Barcelona city metropolitan area and Catalan rural areas}},\ }\href@noop {} {\bibfield  {journal} {\bibinfo  {journal} {Journal of Environmental Sciences}\ }\textbf {\bibinfo {volume} {20}},\ \bibinfo {pages} {1063} (\bibinfo {year} {2008})}\BibitemShut {NoStop}%
\bibitem [{\citenamefont {Mondal}\ \emph {et~al.}(2021)\citenamefont {Mondal}, \citenamefont {Misra}, \citenamefont {Chowdhury}, \citenamefont {Mandal}, \citenamefont {Dutta}, \citenamefont {Baildya}, \citenamefont {Khan}, \citenamefont {Mandal}, \citenamefont {Reza}, \citenamefont {Ahmed} \emph {et~al.}}]{mondal2021exhaled}%
  \BibitemOpen
  \bibfield  {author} {\bibinfo {author} {\bibfnamefont {P.}~\bibnamefont {Mondal}}, \bibinfo {author} {\bibfnamefont {D.}~\bibnamefont {Misra}}, \bibinfo {author} {\bibfnamefont {S.~K.}\ \bibnamefont {Chowdhury}}, \bibinfo {author} {\bibfnamefont {V.}~\bibnamefont {Mandal}}, \bibinfo {author} {\bibfnamefont {T.}~\bibnamefont {Dutta}}, \bibinfo {author} {\bibfnamefont {N.}~\bibnamefont {Baildya}}, \bibinfo {author} {\bibfnamefont {A.~A.}\ \bibnamefont {Khan}}, \bibinfo {author} {\bibfnamefont {M.}~\bibnamefont {Mandal}}, \bibinfo {author} {\bibfnamefont {R.}~\bibnamefont {Reza}}, \bibinfo {author} {\bibfnamefont {M.}~\bibnamefont {Ahmed}}, \emph {et~al.},\ }\bibfield  {title} {\bibinfo {title} {{Exhaled Volatile Organic Compounds (VOCs): a potential biomarkers for chronic disease diagnosis}},\ }\href@noop {} {\bibfield  {journal} {\bibinfo  {journal} {Volatile Org. Compd}\ }\textbf {\bibinfo {volume} {50083}} (\bibinfo {year} {2021})}\BibitemShut {NoStop}%
\bibitem [{\citenamefont {Sethi}\ \emph {et~al.}(2013)\citenamefont {Sethi}, \citenamefont {Nanda},\ and\ \citenamefont {Chakraborty}}]{sethi2013clinical}%
  \BibitemOpen
  \bibfield  {author} {\bibinfo {author} {\bibfnamefont {S.}~\bibnamefont {Sethi}}, \bibinfo {author} {\bibfnamefont {R.}~\bibnamefont {Nanda}},\ and\ \bibinfo {author} {\bibfnamefont {T.}~\bibnamefont {Chakraborty}},\ }\bibfield  {title} {\bibinfo {title} {Clinical application of volatile organic compound analysis for detecting infectious diseases},\ }\href@noop {} {\bibfield  {journal} {\bibinfo  {journal} {Clinical microbiology reviews}\ }\textbf {\bibinfo {volume} {26}},\ \bibinfo {pages} {462} (\bibinfo {year} {2013})}\BibitemShut {NoStop}%
\bibitem [{\citenamefont {Liedberg}\ \emph {et~al.}(1995)\citenamefont {Liedberg}, \citenamefont {Nylander},\ and\ \citenamefont {Lundstr{\"o}m}}]{liedberg1995biosensing}%
  \BibitemOpen
  \bibfield  {author} {\bibinfo {author} {\bibfnamefont {B.}~\bibnamefont {Liedberg}}, \bibinfo {author} {\bibfnamefont {C.}~\bibnamefont {Nylander}},\ and\ \bibinfo {author} {\bibfnamefont {I.}~\bibnamefont {Lundstr{\"o}m}},\ }\bibfield  {title} {\bibinfo {title} {{Biosensing with surface plasmon resonance-how it all started}},\ }\href@noop {} {\bibfield  {journal} {\bibinfo  {journal} {Biosensors and Bioelectronics}\ }\textbf {\bibinfo {volume} {10}},\ \bibinfo {pages} {i} (\bibinfo {year} {1995})}\BibitemShut {NoStop}%
\bibitem [{\citenamefont {Liedberg}\ \emph {et~al.}(1983)\citenamefont {Liedberg}, \citenamefont {Nylander},\ and\ \citenamefont {Lunstr{\"o}m}}]{liedberg1983surface}%
  \BibitemOpen
  \bibfield  {author} {\bibinfo {author} {\bibfnamefont {B.}~\bibnamefont {Liedberg}}, \bibinfo {author} {\bibfnamefont {C.}~\bibnamefont {Nylander}},\ and\ \bibinfo {author} {\bibfnamefont {I.}~\bibnamefont {Lunstr{\"o}m}},\ }\bibfield  {title} {\bibinfo {title} {Surface plasmon resonance for gas detection and biosensing},\ }\href@noop {} {\bibfield  {journal} {\bibinfo  {journal} {Sensors and actuators}\ }\textbf {\bibinfo {volume} {4}},\ \bibinfo {pages} {299} (\bibinfo {year} {1983})}\BibitemShut {NoStop}%
\bibitem [{\citenamefont {Singh}(2016)}]{singh2016spr}%
  \BibitemOpen
  \bibfield  {author} {\bibinfo {author} {\bibfnamefont {P.}~\bibnamefont {Singh}},\ }\bibfield  {title} {\bibinfo {title} {{SPR biosensors: historical perspectives and current challenges}},\ }\href@noop {} {\bibfield  {journal} {\bibinfo  {journal} {Sensors and actuators B: Chemical}\ }\textbf {\bibinfo {volume} {229}},\ \bibinfo {pages} {110} (\bibinfo {year} {2016})}\BibitemShut {NoStop}%
\bibitem [{\citenamefont {Fabricius}\ \emph {et~al.}(1992)\citenamefont {Fabricius}, \citenamefont {Gauglitz},\ and\ \citenamefont {Ingenhoff}}]{fabricius1992gas}%
  \BibitemOpen
  \bibfield  {author} {\bibinfo {author} {\bibfnamefont {N.}~\bibnamefont {Fabricius}}, \bibinfo {author} {\bibfnamefont {G.}~\bibnamefont {Gauglitz}},\ and\ \bibinfo {author} {\bibfnamefont {J.}~\bibnamefont {Ingenhoff}},\ }\bibfield  {title} {\bibinfo {title} {{A gas sensor based on an integrated optical Mach-Zehnder interferometer}},\ }\href@noop {} {\bibfield  {journal} {\bibinfo  {journal} {Sensors and Actuators B: Chemical}\ }\textbf {\bibinfo {volume} {7}},\ \bibinfo {pages} {672} (\bibinfo {year} {1992})}\BibitemShut {NoStop}%
\bibitem [{\citenamefont {Hao}\ and\ \citenamefont {Chiang}(2017)}]{hao2017graphene}%
  \BibitemOpen
  \bibfield  {author} {\bibinfo {author} {\bibfnamefont {T.}~\bibnamefont {Hao}}\ and\ \bibinfo {author} {\bibfnamefont {K.~S.}\ \bibnamefont {Chiang}},\ }\bibfield  {title} {\bibinfo {title} {{Graphene-based ammonia-gas sensor using in-fiber Mach-Zehnder interferometer}},\ }\href@noop {} {\bibfield  {journal} {\bibinfo  {journal} {IEEE Photonics Technology Letters}\ }\textbf {\bibinfo {volume} {29}},\ \bibinfo {pages} {2035} (\bibinfo {year} {2017})}\BibitemShut {NoStop}%
\bibitem [{\citenamefont {Schubert}\ \emph {et~al.}(1997)\citenamefont {Schubert}, \citenamefont {Haase}, \citenamefont {K{\"u}ck},\ and\ \citenamefont {Gottfried-Gottfried}}]{schubert1997refractive}%
  \BibitemOpen
  \bibfield  {author} {\bibinfo {author} {\bibfnamefont {T.}~\bibnamefont {Schubert}}, \bibinfo {author} {\bibfnamefont {N.}~\bibnamefont {Haase}}, \bibinfo {author} {\bibfnamefont {H.}~\bibnamefont {K{\"u}ck}},\ and\ \bibinfo {author} {\bibfnamefont {R.}~\bibnamefont {Gottfried-Gottfried}},\ }\bibfield  {title} {\bibinfo {title} {{Refractive-index measurements using an integrated Mach-Zehnder interferometer}},\ }\href@noop {} {\bibfield  {journal} {\bibinfo  {journal} {Sensors and Actuators A: Physical}\ }\textbf {\bibinfo {volume} {60}},\ \bibinfo {pages} {108} (\bibinfo {year} {1997})}\BibitemShut {NoStop}%
\bibitem [{\citenamefont {Liu}\ \emph {et~al.}(2019)\citenamefont {Liu}, \citenamefont {Wang}, \citenamefont {Biswas}, \citenamefont {Palit}, \citenamefont {Zhou},\ and\ \citenamefont {Sun}}]{liu2019optofluidic}%
  \BibitemOpen
  \bibfield  {author} {\bibinfo {author} {\bibfnamefont {Y.}~\bibnamefont {Liu}}, \bibinfo {author} {\bibfnamefont {S.}~\bibnamefont {Wang}}, \bibinfo {author} {\bibfnamefont {P.}~\bibnamefont {Biswas}}, \bibinfo {author} {\bibfnamefont {P.}~\bibnamefont {Palit}}, \bibinfo {author} {\bibfnamefont {W.}~\bibnamefont {Zhou}},\ and\ \bibinfo {author} {\bibfnamefont {Y.}~\bibnamefont {Sun}},\ }\bibfield  {title} {\bibinfo {title} {{Optofluidic vapor sensing with free-space coupled 2D photonic crystal slabs}},\ }\href@noop {} {\bibfield  {journal} {\bibinfo  {journal} {Scientific Reports}\ }\textbf {\bibinfo {volume} {9}},\ \bibinfo {pages} {1} (\bibinfo {year} {2019})}\BibitemShut {NoStop}%
\bibitem [{\citenamefont {Drayton}\ \emph {et~al.}(2019)\citenamefont {Drayton}, \citenamefont {Barth},\ and\ \citenamefont {Krauss}}]{drayton2019guided}%
  \BibitemOpen
  \bibfield  {author} {\bibinfo {author} {\bibfnamefont {A.}~\bibnamefont {Drayton}}, \bibinfo {author} {\bibfnamefont {I.}~\bibnamefont {Barth}},\ and\ \bibinfo {author} {\bibfnamefont {T.~F.}\ \bibnamefont {Krauss}},\ }\bibfield  {title} {\bibinfo {title} {Guided mode resonances and photonic crystals for biosensing and imaging},\ }in\ \href@noop {} {\emph {\bibinfo {booktitle} {Semiconductors and Semimetals}}},\ Vol.\ \bibinfo {volume} {100}\ (\bibinfo  {publisher} {Elsevier},\ \bibinfo {year} {2019})\ pp.\ \bibinfo {pages} {115--148}\BibitemShut {NoStop}%
\bibitem [{\citenamefont {Pitruzzello}\ and\ \citenamefont {Krauss}(2018)}]{pitruzzello2018photonic}%
  \BibitemOpen
  \bibfield  {author} {\bibinfo {author} {\bibfnamefont {G.}~\bibnamefont {Pitruzzello}}\ and\ \bibinfo {author} {\bibfnamefont {T.~F.}\ \bibnamefont {Krauss}},\ }\bibfield  {title} {\bibinfo {title} {Photonic crystal resonances for sensing and imaging},\ }\href@noop {} {\bibfield  {journal} {\bibinfo  {journal} {Journal of Optics}\ }\textbf {\bibinfo {volume} {20}},\ \bibinfo {pages} {073004} (\bibinfo {year} {2018})}\BibitemShut {NoStop}%
\bibitem [{\citenamefont {Barth}\ \emph {et~al.}(2020)\citenamefont {Barth}, \citenamefont {Conteduca}, \citenamefont {Reardon}, \citenamefont {Johnson},\ and\ \citenamefont {Krauss}}]{barth2020common}%
  \BibitemOpen
  \bibfield  {author} {\bibinfo {author} {\bibfnamefont {I.}~\bibnamefont {Barth}}, \bibinfo {author} {\bibfnamefont {D.}~\bibnamefont {Conteduca}}, \bibinfo {author} {\bibfnamefont {C.}~\bibnamefont {Reardon}}, \bibinfo {author} {\bibfnamefont {S.}~\bibnamefont {Johnson}},\ and\ \bibinfo {author} {\bibfnamefont {T.~F.}\ \bibnamefont {Krauss}},\ }\bibfield  {title} {\bibinfo {title} {Common-path interferometric label-free protein sensing with resonant dielectric nanostructures},\ }\href@noop {} {\bibfield  {journal} {\bibinfo  {journal} {Light: Science \& Applications}\ }\textbf {\bibinfo {volume} {9}},\ \bibinfo {pages} {1} (\bibinfo {year} {2020})}\BibitemShut {NoStop}%
\bibitem [{\citenamefont {Grigorenko}\ \emph {et~al.}(1999)\citenamefont {Grigorenko}, \citenamefont {Nikitin},\ and\ \citenamefont {Kabashin}}]{grigorenko1999phase}%
  \BibitemOpen
  \bibfield  {author} {\bibinfo {author} {\bibfnamefont {A.}~\bibnamefont {Grigorenko}}, \bibinfo {author} {\bibfnamefont {P.}~\bibnamefont {Nikitin}},\ and\ \bibinfo {author} {\bibfnamefont {A.}~\bibnamefont {Kabashin}},\ }\bibfield  {title} {\bibinfo {title} {Phase jumps and interferometric surface plasmon resonance imaging},\ }\href@noop {} {\bibfield  {journal} {\bibinfo  {journal} {Applied Physics Letters}\ }\textbf {\bibinfo {volume} {75}},\ \bibinfo {pages} {3917} (\bibinfo {year} {1999})}\BibitemShut {NoStop}%
\bibitem [{\citenamefont {Kabashin}\ and\ \citenamefont {Nikitin}(1997)}]{kabashin1997interferometer}%
  \BibitemOpen
  \bibfield  {author} {\bibinfo {author} {\bibfnamefont {A.~V.}\ \bibnamefont {Kabashin}}\ and\ \bibinfo {author} {\bibfnamefont {P.~I.}\ \bibnamefont {Nikitin}},\ }\bibfield  {title} {\bibinfo {title} {Interferometer based on a surface-plasmon resonance for sensor applications},\ }\href@noop {} {\bibfield  {journal} {\bibinfo  {journal} {Quantum Electronics}\ }\textbf {\bibinfo {volume} {27}},\ \bibinfo {pages} {653} (\bibinfo {year} {1997})}\BibitemShut {NoStop}%
\bibitem [{\citenamefont {Kabashin}\ \emph {et~al.}(2009)\citenamefont {Kabashin}, \citenamefont {Patskovsky},\ and\ \citenamefont {Grigorenko}}]{kabashin2009phase}%
  \BibitemOpen
  \bibfield  {author} {\bibinfo {author} {\bibfnamefont {A.~V.}\ \bibnamefont {Kabashin}}, \bibinfo {author} {\bibfnamefont {S.}~\bibnamefont {Patskovsky}},\ and\ \bibinfo {author} {\bibfnamefont {A.~N.}\ \bibnamefont {Grigorenko}},\ }\bibfield  {title} {\bibinfo {title} {Phase and amplitude sensitivities in surface plasmon resonance bio and chemical sensing},\ }\href@noop {} {\bibfield  {journal} {\bibinfo  {journal} {Optics Express}\ }\textbf {\bibinfo {volume} {17}},\ \bibinfo {pages} {21191} (\bibinfo {year} {2009})}\BibitemShut {NoStop}%
\bibitem [{\citenamefont {Boriskina}\ and\ \citenamefont {Tsurimaki}(2018)}]{boriskina2018sensitive}%
  \BibitemOpen
  \bibfield  {author} {\bibinfo {author} {\bibfnamefont {S.~V.}\ \bibnamefont {Boriskina}}\ and\ \bibinfo {author} {\bibfnamefont {Y.}~\bibnamefont {Tsurimaki}},\ }\bibfield  {title} {\bibinfo {title} {{Sensitive singular-phase optical detection without phase measurements with Tamm plasmons}},\ }\href@noop {} {\bibfield  {journal} {\bibinfo  {journal} {Journal of Physics: Condensed Matter}\ }\textbf {\bibinfo {volume} {30}},\ \bibinfo {pages} {224003} (\bibinfo {year} {2018})}\BibitemShut {NoStop}%
\bibitem [{\citenamefont {Liu}\ \emph {et~al.}(2023)\citenamefont {Liu}, \citenamefont {Chen}, \citenamefont {Hu}, \citenamefont {Fan}, \citenamefont {Christodoulides}, \citenamefont {Zhao},\ and\ \citenamefont {Qiu}}]{liu2023}%
  \BibitemOpen
  \bibfield  {author} {\bibinfo {author} {\bibfnamefont {M.}~\bibnamefont {Liu}}, \bibinfo {author} {\bibfnamefont {W.}~\bibnamefont {Chen}}, \bibinfo {author} {\bibfnamefont {G.}~\bibnamefont {Hu}}, \bibinfo {author} {\bibfnamefont {S.}~\bibnamefont {Fan}}, \bibinfo {author} {\bibfnamefont {D.~N.}\ \bibnamefont {Christodoulides}}, \bibinfo {author} {\bibfnamefont {C.}~\bibnamefont {Zhao}},\ and\ \bibinfo {author} {\bibfnamefont {C.-W.}\ \bibnamefont {Qiu}},\ }\bibfield  {title} {\bibinfo {title} {Spectral phase singularity and topological behavior in perfect absorption},\ }\href@noop {} {\bibfield  {journal} {\bibinfo  {journal} {Physical Review B}\ }\textbf {\bibinfo {volume} {107}},\ \bibinfo {pages} {L241403} (\bibinfo {year} {2023})}\BibitemShut {NoStop}%
\bibitem [{\citenamefont {Song}\ \emph {et~al.}(2017)\citenamefont {Song}, \citenamefont {Zhang}, \citenamefont {Duan}, \citenamefont {Liu}, \citenamefont {Gao}, \citenamefont {Singer}, \citenamefont {Ji}, \citenamefont {Cheney}, \citenamefont {Zeng}, \citenamefont {Chen} \emph {et~al.}}]{song2017dispersion}%
  \BibitemOpen
  \bibfield  {author} {\bibinfo {author} {\bibfnamefont {H.}~\bibnamefont {Song}}, \bibinfo {author} {\bibfnamefont {N.}~\bibnamefont {Zhang}}, \bibinfo {author} {\bibfnamefont {J.}~\bibnamefont {Duan}}, \bibinfo {author} {\bibfnamefont {Z.}~\bibnamefont {Liu}}, \bibinfo {author} {\bibfnamefont {J.}~\bibnamefont {Gao}}, \bibinfo {author} {\bibfnamefont {M.~H.}\ \bibnamefont {Singer}}, \bibinfo {author} {\bibfnamefont {D.}~\bibnamefont {Ji}}, \bibinfo {author} {\bibfnamefont {A.~R.}\ \bibnamefont {Cheney}}, \bibinfo {author} {\bibfnamefont {X.}~\bibnamefont {Zeng}}, \bibinfo {author} {\bibfnamefont {B.}~\bibnamefont {Chen}}, \emph {et~al.},\ }\bibfield  {title} {\bibinfo {title} {Dispersion topological darkness at multiple wavelengths and polarization states},\ }\href@noop {} {\bibfield  {journal} {\bibinfo  {journal} {Advanced Optical Materials}\ }\textbf {\bibinfo {volume} {5}},\ \bibinfo {pages} {1700166} (\bibinfo {year} {2017})}\BibitemShut {NoStop}%
\bibitem [{\citenamefont {Kravets}\ \emph {et~al.}(2013)\citenamefont {Kravets}, \citenamefont {Schedin}, \citenamefont {Jalil}, \citenamefont {Britnell}, \citenamefont {Gorbachev}, \citenamefont {Ansell}, \citenamefont {Thackray}, \citenamefont {Novoselov}, \citenamefont {Geim}, \citenamefont {Kabashin} \emph {et~al.}}]{kravets2013singular}%
  \BibitemOpen
  \bibfield  {author} {\bibinfo {author} {\bibfnamefont {V.}~\bibnamefont {Kravets}}, \bibinfo {author} {\bibfnamefont {F.}~\bibnamefont {Schedin}}, \bibinfo {author} {\bibfnamefont {R.}~\bibnamefont {Jalil}}, \bibinfo {author} {\bibfnamefont {L.}~\bibnamefont {Britnell}}, \bibinfo {author} {\bibfnamefont {R.}~\bibnamefont {Gorbachev}}, \bibinfo {author} {\bibfnamefont {D.}~\bibnamefont {Ansell}}, \bibinfo {author} {\bibfnamefont {B.}~\bibnamefont {Thackray}}, \bibinfo {author} {\bibfnamefont {K.}~\bibnamefont {Novoselov}}, \bibinfo {author} {\bibfnamefont {A.}~\bibnamefont {Geim}}, \bibinfo {author} {\bibfnamefont {A.~V.}\ \bibnamefont {Kabashin}}, \emph {et~al.},\ }\bibfield  {title} {\bibinfo {title} {Singular phase nano-optics in plasmonic metamaterials for label-free single-molecule detection},\ }\href@noop {} {\bibfield  {journal} {\bibinfo  {journal} {Nature materials}\ }\textbf {\bibinfo {volume} {12}},\ \bibinfo {pages} {304} (\bibinfo {year} {2013})}\BibitemShut {NoStop}%
\bibitem [{\citenamefont {Sreekanth}\ \emph {et~al.}(2018{\natexlab{a}})\citenamefont {Sreekanth}, \citenamefont {Sreejith}, \citenamefont {Han}, \citenamefont {Mishra}, \citenamefont {Chen}, \citenamefont {Sun}, \citenamefont {Lim},\ and\ \citenamefont {Singh}}]{sreekanth2018biosensing}%
  \BibitemOpen
  \bibfield  {author} {\bibinfo {author} {\bibfnamefont {K.~V.}\ \bibnamefont {Sreekanth}}, \bibinfo {author} {\bibfnamefont {S.}~\bibnamefont {Sreejith}}, \bibinfo {author} {\bibfnamefont {S.}~\bibnamefont {Han}}, \bibinfo {author} {\bibfnamefont {A.}~\bibnamefont {Mishra}}, \bibinfo {author} {\bibfnamefont {X.}~\bibnamefont {Chen}}, \bibinfo {author} {\bibfnamefont {H.}~\bibnamefont {Sun}}, \bibinfo {author} {\bibfnamefont {C.~T.}\ \bibnamefont {Lim}},\ and\ \bibinfo {author} {\bibfnamefont {R.}~\bibnamefont {Singh}},\ }\bibfield  {title} {\bibinfo {title} {Biosensing with the singular phase of an ultrathin metal-dielectric nanophotonic cavity},\ }\href@noop {} {\bibfield  {journal} {\bibinfo  {journal} {Nature communications}\ }\textbf {\bibinfo {volume} {9}},\ \bibinfo {pages} {1} (\bibinfo {year} {2018}{\natexlab{a}})}\BibitemShut {NoStop}%
\bibitem [{\citenamefont {Cueff}\ \emph {et~al.}(2021)\citenamefont {Cueff}, \citenamefont {Taute}, \citenamefont {Bourgade}, \citenamefont {Lumeau}, \citenamefont {Monfray}, \citenamefont {Song}, \citenamefont {Genevet}, \citenamefont {Devif}, \citenamefont {Letartre},\ and\ \citenamefont {Berguiga}}]{cueff2021reconfigurable}%
  \BibitemOpen
  \bibfield  {author} {\bibinfo {author} {\bibfnamefont {S.}~\bibnamefont {Cueff}}, \bibinfo {author} {\bibfnamefont {A.}~\bibnamefont {Taute}}, \bibinfo {author} {\bibfnamefont {A.}~\bibnamefont {Bourgade}}, \bibinfo {author} {\bibfnamefont {J.}~\bibnamefont {Lumeau}}, \bibinfo {author} {\bibfnamefont {S.}~\bibnamefont {Monfray}}, \bibinfo {author} {\bibfnamefont {Q.}~\bibnamefont {Song}}, \bibinfo {author} {\bibfnamefont {P.}~\bibnamefont {Genevet}}, \bibinfo {author} {\bibfnamefont {B.}~\bibnamefont {Devif}}, \bibinfo {author} {\bibfnamefont {X.}~\bibnamefont {Letartre}},\ and\ \bibinfo {author} {\bibfnamefont {L.}~\bibnamefont {Berguiga}},\ }\bibfield  {title} {\bibinfo {title} {Reconfigurable flat optics with programmable reflection amplitude using lithography-free phase-change material ultra-thin films},\ }\href@noop {} {\bibfield  {journal} {\bibinfo  {journal} {Advanced Optical Materials}\ }\textbf {\bibinfo {volume} {9}},\ \bibinfo {pages} {2001291} (\bibinfo {year} {2021})}\BibitemShut {NoStop}%
\bibitem [{\citenamefont {Sreekanth}\ \emph {et~al.}(2018{\natexlab{b}})\citenamefont {Sreekanth}, \citenamefont {Han},\ and\ \citenamefont {Singh}}]{sreekanth2018ge2sb2te5}%
  \BibitemOpen
  \bibfield  {author} {\bibinfo {author} {\bibfnamefont {K.~V.}\ \bibnamefont {Sreekanth}}, \bibinfo {author} {\bibfnamefont {S.}~\bibnamefont {Han}},\ and\ \bibinfo {author} {\bibfnamefont {R.}~\bibnamefont {Singh}},\ }\bibfield  {title} {\bibinfo {title} {{Ge$_2$Sb$_2$Te$_5$-based tunable perfect absorber cavity with phase singularity at visible frequencies}},\ }\href@noop {} {\bibfield  {journal} {\bibinfo  {journal} {Advanced Materials}\ }\textbf {\bibinfo {volume} {30}},\ \bibinfo {pages} {1706696} (\bibinfo {year} {2018}{\natexlab{b}})}\BibitemShut {NoStop}%
\bibitem [{\citenamefont {Piper}\ \emph {et~al.}(2014)\citenamefont {Piper}, \citenamefont {Liu},\ and\ \citenamefont {Fan}}]{piper2014total}%
  \BibitemOpen
  \bibfield  {author} {\bibinfo {author} {\bibfnamefont {J.~R.}\ \bibnamefont {Piper}}, \bibinfo {author} {\bibfnamefont {V.}~\bibnamefont {Liu}},\ and\ \bibinfo {author} {\bibfnamefont {S.}~\bibnamefont {Fan}},\ }\bibfield  {title} {\bibinfo {title} {Total absorption by degenerate critical coupling},\ }\href@noop {} {\bibfield  {journal} {\bibinfo  {journal} {Applied Physics Letters}\ }\textbf {\bibinfo {volume} {104}},\ \bibinfo {pages} {251110} (\bibinfo {year} {2014})}\BibitemShut {NoStop}%
\bibitem [{\citenamefont {Berguiga}\ \emph {et~al.}(2021)\citenamefont {Berguiga}, \citenamefont {Ferrier}, \citenamefont {Jamois}, \citenamefont {Benyattou}, \citenamefont {Letartre},\ and\ \citenamefont {Cueff}}]{berguiga2021ultimate}%
  \BibitemOpen
  \bibfield  {author} {\bibinfo {author} {\bibfnamefont {L.}~\bibnamefont {Berguiga}}, \bibinfo {author} {\bibfnamefont {L.}~\bibnamefont {Ferrier}}, \bibinfo {author} {\bibfnamefont {C.}~\bibnamefont {Jamois}}, \bibinfo {author} {\bibfnamefont {T.}~\bibnamefont {Benyattou}}, \bibinfo {author} {\bibfnamefont {X.}~\bibnamefont {Letartre}},\ and\ \bibinfo {author} {\bibfnamefont {S.}~\bibnamefont {Cueff}},\ }\bibfield  {title} {\bibinfo {title} {Ultimate phase sensitivity in surface plasmon resonance sensors by tuning critical coupling with phase change materials},\ }\href@noop {} {\bibfield  {journal} {\bibinfo  {journal} {Optics Express}\ }\textbf {\bibinfo {volume} {29}},\ \bibinfo {pages} {42162} (\bibinfo {year} {2021})}\BibitemShut {NoStop}%
\bibitem [{\citenamefont {Kim}\ \emph {et~al.}(2022)\citenamefont {Kim}, \citenamefont {Ko}, \citenamefont {Yoo}, \citenamefont {Kim}, \citenamefont {Lee}, \citenamefont {Ishii},\ and\ \citenamefont {Song}}]{kim2022single}%
  \BibitemOpen
  \bibfield  {author} {\bibinfo {author} {\bibfnamefont {S.~H.}\ \bibnamefont {Kim}}, \bibinfo {author} {\bibfnamefont {J.~H.}\ \bibnamefont {Ko}}, \bibinfo {author} {\bibfnamefont {Y.~J.}\ \bibnamefont {Yoo}}, \bibinfo {author} {\bibfnamefont {M.~S.}\ \bibnamefont {Kim}}, \bibinfo {author} {\bibfnamefont {G.~J.}\ \bibnamefont {Lee}}, \bibinfo {author} {\bibfnamefont {S.}~\bibnamefont {Ishii}},\ and\ \bibinfo {author} {\bibfnamefont {Y.~M.}\ \bibnamefont {Song}},\ }\bibfield  {title} {\bibinfo {title} {{Single-material, near-infrared selective absorber based on refractive index-tunable Tamm plasmon structure}},\ }\href@noop {} {\bibfield  {journal} {\bibinfo  {journal} {Advanced Optical Materials}\ }\textbf {\bibinfo {volume} {10}},\ \bibinfo {pages} {2102388} (\bibinfo {year} {2022})}\BibitemShut {NoStop}%
\bibitem [{\citenamefont {Ma}\ \emph {et~al.}(2023)\citenamefont {Ma}, \citenamefont {Shen}, \citenamefont {Sanchez}, \citenamefont {Yu}, \citenamefont {Wang}, \citenamefont {Sun}, \citenamefont {Wang},\ and\ \citenamefont {Hu}}]{ma2023excitons}%
  \BibitemOpen
  \bibfield  {author} {\bibinfo {author} {\bibfnamefont {G.}~\bibnamefont {Ma}}, \bibinfo {author} {\bibfnamefont {W.}~\bibnamefont {Shen}}, \bibinfo {author} {\bibfnamefont {D.~S.}\ \bibnamefont {Sanchez}}, \bibinfo {author} {\bibfnamefont {Y.}~\bibnamefont {Yu}}, \bibinfo {author} {\bibfnamefont {H.}~\bibnamefont {Wang}}, \bibinfo {author} {\bibfnamefont {L.}~\bibnamefont {Sun}}, \bibinfo {author} {\bibfnamefont {X.}~\bibnamefont {Wang}},\ and\ \bibinfo {author} {\bibfnamefont {C.}~\bibnamefont {Hu}},\ }\bibfield  {title} {\bibinfo {title} {Excitons enabled topological phase singularity in a single atomic layer},\ }\href@noop {} {\bibfield  {journal} {\bibinfo  {journal} {ACS nano}\ }\textbf {\bibinfo {volume} {17}},\ \bibinfo {pages} {17751} (\bibinfo {year} {2023})}\BibitemShut {NoStop}%
\bibitem [{\citenamefont {Ermolaev}\ \emph {et~al.}(2022)\citenamefont {Ermolaev}, \citenamefont {Voronin}, \citenamefont {Baranov}, \citenamefont {Kravets}, \citenamefont {Tselikov}, \citenamefont {Stebunov}, \citenamefont {Yakubovsky}, \citenamefont {Novikov}, \citenamefont {Vyshnevyy}, \citenamefont {Mazitov} \emph {et~al.}}]{ermolaev2022topological}%
  \BibitemOpen
  \bibfield  {author} {\bibinfo {author} {\bibfnamefont {G.}~\bibnamefont {Ermolaev}}, \bibinfo {author} {\bibfnamefont {K.}~\bibnamefont {Voronin}}, \bibinfo {author} {\bibfnamefont {D.~G.}\ \bibnamefont {Baranov}}, \bibinfo {author} {\bibfnamefont {V.}~\bibnamefont {Kravets}}, \bibinfo {author} {\bibfnamefont {G.}~\bibnamefont {Tselikov}}, \bibinfo {author} {\bibfnamefont {Y.}~\bibnamefont {Stebunov}}, \bibinfo {author} {\bibfnamefont {D.}~\bibnamefont {Yakubovsky}}, \bibinfo {author} {\bibfnamefont {S.}~\bibnamefont {Novikov}}, \bibinfo {author} {\bibfnamefont {A.}~\bibnamefont {Vyshnevyy}}, \bibinfo {author} {\bibfnamefont {A.}~\bibnamefont {Mazitov}}, \emph {et~al.},\ }\bibfield  {title} {\bibinfo {title} {Topological phase singularities in atomically thin high-refractive-index materials},\ }\href@noop {} {\bibfield  {journal} {\bibinfo  {journal} {Nature communications}\ }\textbf {\bibinfo {volume} {13}},\ \bibinfo {pages} {2049} (\bibinfo {year} {2022})}\BibitemShut {NoStop}%
\bibitem [{\citenamefont {Mkhitaryan}\ \emph {et~al.}(2017)\citenamefont {Mkhitaryan}, \citenamefont {Ghosh}, \citenamefont {Rud{\'e}}, \citenamefont {Canet-Ferrer}, \citenamefont {Maniyara}, \citenamefont {Gopalan},\ and\ \citenamefont {Pruneri}}]{mkhitaryan2017tunable}%
  \BibitemOpen
  \bibfield  {author} {\bibinfo {author} {\bibfnamefont {V.~K.}\ \bibnamefont {Mkhitaryan}}, \bibinfo {author} {\bibfnamefont {D.~S.}\ \bibnamefont {Ghosh}}, \bibinfo {author} {\bibfnamefont {M.}~\bibnamefont {Rud{\'e}}}, \bibinfo {author} {\bibfnamefont {J.}~\bibnamefont {Canet-Ferrer}}, \bibinfo {author} {\bibfnamefont {R.~A.}\ \bibnamefont {Maniyara}}, \bibinfo {author} {\bibfnamefont {K.~K.}\ \bibnamefont {Gopalan}},\ and\ \bibinfo {author} {\bibfnamefont {V.}~\bibnamefont {Pruneri}},\ }\bibfield  {title} {\bibinfo {title} {{Tunable complete optical absorption in multilayer structures including Ge$_2$Sb$_2$Te$_5$ without lithographic patterns}},\ }\href@noop {} {\bibfield  {journal} {\bibinfo  {journal} {Advanced Optical Materials}\ }\textbf {\bibinfo {volume} {5}},\ \bibinfo {pages} {1600452} (\bibinfo {year} {2017})}\BibitemShut {NoStop}%
\bibitem [{\citenamefont {Yue}\ \emph {et~al.}(2023)\citenamefont {Yue}, \citenamefont {Wang}, \citenamefont {Yan}, \citenamefont {Wang}, \citenamefont {Wang}, \citenamefont {Wang}, \citenamefont {Zhang},\ and\ \citenamefont {Wang}}]{yue2023high}%
  \BibitemOpen
  \bibfield  {author} {\bibinfo {author} {\bibfnamefont {X.}~\bibnamefont {Yue}}, \bibinfo {author} {\bibfnamefont {T.}~\bibnamefont {Wang}}, \bibinfo {author} {\bibfnamefont {R.}~\bibnamefont {Yan}}, \bibinfo {author} {\bibfnamefont {L.}~\bibnamefont {Wang}}, \bibinfo {author} {\bibfnamefont {H.}~\bibnamefont {Wang}}, \bibinfo {author} {\bibfnamefont {Y.}~\bibnamefont {Wang}}, \bibinfo {author} {\bibfnamefont {J.}~\bibnamefont {Zhang}},\ and\ \bibinfo {author} {\bibfnamefont {J.}~\bibnamefont {Wang}},\ }\bibfield  {title} {\bibinfo {title} {{High-sensitivity refractive index sensing with the singular phase in normal incidence of an asymmetric Fabry--Perot cavity modulated by grating}},\ }\href@noop {} {\bibfield  {journal} {\bibinfo  {journal} {Optics \& Laser Technology}\ }\textbf {\bibinfo {volume} {157}},\ \bibinfo {pages} {108697} (\bibinfo {year} {2023})}\BibitemShut {NoStop}%
\bibitem [{\citenamefont {Tsurimaki}\ \emph {et~al.}(2018)\citenamefont {Tsurimaki}, \citenamefont {Tong}, \citenamefont {Boriskina}, \citenamefont {Semenov}, \citenamefont {Ayzatsky}, \citenamefont {Machekhin}, \citenamefont {Chen},\ and\ \citenamefont {Boriskina}}]{tsurimaki2018topological}%
  \BibitemOpen
  \bibfield  {author} {\bibinfo {author} {\bibfnamefont {Y.}~\bibnamefont {Tsurimaki}}, \bibinfo {author} {\bibfnamefont {J.~K.}\ \bibnamefont {Tong}}, \bibinfo {author} {\bibfnamefont {V.~N.}\ \bibnamefont {Boriskina}}, \bibinfo {author} {\bibfnamefont {A.}~\bibnamefont {Semenov}}, \bibinfo {author} {\bibfnamefont {M.~I.}\ \bibnamefont {Ayzatsky}}, \bibinfo {author} {\bibfnamefont {Y.~P.}\ \bibnamefont {Machekhin}}, \bibinfo {author} {\bibfnamefont {G.}~\bibnamefont {Chen}},\ and\ \bibinfo {author} {\bibfnamefont {S.~V.}\ \bibnamefont {Boriskina}},\ }\bibfield  {title} {\bibinfo {title} {{Topological engineering of interfacial optical Tamm states for highly sensitive near-singular-phase optical detection}},\ }\href@noop {} {\bibfield  {journal} {\bibinfo  {journal} {ACS Photonics}\ }\textbf {\bibinfo {volume} {5}},\ \bibinfo {pages} {929} (\bibinfo {year} {2018})}\BibitemShut {NoStop}%
\bibitem [{\citenamefont {Sakotic}\ \emph {et~al.}(2021)\citenamefont {Sakotic}, \citenamefont {Krasnok}, \citenamefont {Al{\'u}},\ and\ \citenamefont {Jankovic}}]{sakotic2021topological}%
  \BibitemOpen
  \bibfield  {author} {\bibinfo {author} {\bibfnamefont {Z.}~\bibnamefont {Sakotic}}, \bibinfo {author} {\bibfnamefont {A.}~\bibnamefont {Krasnok}}, \bibinfo {author} {\bibfnamefont {A.}~\bibnamefont {Al{\'u}}},\ and\ \bibinfo {author} {\bibfnamefont {N.}~\bibnamefont {Jankovic}},\ }\bibfield  {title} {\bibinfo {title} {Topological scattering singularities and embedded eigenstates for polarization control and sensing applications},\ }\href@noop {} {\bibfield  {journal} {\bibinfo  {journal} {Photonics Research}\ }\textbf {\bibinfo {volume} {9}},\ \bibinfo {pages} {1310} (\bibinfo {year} {2021})}\BibitemShut {NoStop}%
\bibitem [{\citenamefont {Vasi{\'c}}\ and\ \citenamefont {Gaji{\'c}}(2014)}]{vasic2014enhanced}%
  \BibitemOpen
  \bibfield  {author} {\bibinfo {author} {\bibfnamefont {B.}~\bibnamefont {Vasi{\'c}}}\ and\ \bibinfo {author} {\bibfnamefont {R.}~\bibnamefont {Gaji{\'c}}},\ }\bibfield  {title} {\bibinfo {title} {Enhanced phase sensitivity of metamaterial absorbers near the point of darkness},\ }\href@noop {} {\bibfield  {journal} {\bibinfo  {journal} {Journal of Applied Physics}\ }\textbf {\bibinfo {volume} {116}},\ \bibinfo {pages} {023102} (\bibinfo {year} {2014})}\BibitemShut {NoStop}%
\bibitem [{\citenamefont {Nikitin}\ \emph {et~al.}(1999)\citenamefont {Nikitin}, \citenamefont {Beloglazov}, \citenamefont {Kochergin}, \citenamefont {Valeiko},\ and\ \citenamefont {Ksenevich}}]{nikitin1999surface}%
  \BibitemOpen
  \bibfield  {author} {\bibinfo {author} {\bibfnamefont {P.}~\bibnamefont {Nikitin}}, \bibinfo {author} {\bibfnamefont {A.}~\bibnamefont {Beloglazov}}, \bibinfo {author} {\bibfnamefont {V.}~\bibnamefont {Kochergin}}, \bibinfo {author} {\bibfnamefont {M.}~\bibnamefont {Valeiko}},\ and\ \bibinfo {author} {\bibfnamefont {T.}~\bibnamefont {Ksenevich}},\ }\bibfield  {title} {\bibinfo {title} {Surface plasmon resonance interferometry for biological and chemical sensing},\ }\href@noop {} {\bibfield  {journal} {\bibinfo  {journal} {Sensors and Actuators B: Chemical}\ }\textbf {\bibinfo {volume} {54}},\ \bibinfo {pages} {43} (\bibinfo {year} {1999})}\BibitemShut {NoStop}%
\bibitem [{\citenamefont {Augui{\'e}}\ \emph {et~al.}(2014)\citenamefont {Augui{\'e}}, \citenamefont {Fuertes}, \citenamefont {Angelom{\'e}}, \citenamefont {Abdala}, \citenamefont {Soler~Illia},\ and\ \citenamefont {Fainstein}}]{auguie2014tamm}%
  \BibitemOpen
  \bibfield  {author} {\bibinfo {author} {\bibfnamefont {B.}~\bibnamefont {Augui{\'e}}}, \bibinfo {author} {\bibfnamefont {M.~C.}\ \bibnamefont {Fuertes}}, \bibinfo {author} {\bibfnamefont {P.~C.}\ \bibnamefont {Angelom{\'e}}}, \bibinfo {author} {\bibfnamefont {N.~L.}\ \bibnamefont {Abdala}}, \bibinfo {author} {\bibfnamefont {G.~J.}\ \bibnamefont {Soler~Illia}},\ and\ \bibinfo {author} {\bibfnamefont {A.}~\bibnamefont {Fainstein}},\ }\bibfield  {title} {\bibinfo {title} {Tamm plasmon resonance in mesoporous multilayers: toward a sensing application},\ }\href@noop {} {\bibfield  {journal} {\bibinfo  {journal} {Acs Photonics}\ }\textbf {\bibinfo {volume} {1}},\ \bibinfo {pages} {775} (\bibinfo {year} {2014})}\BibitemShut {NoStop}%
\bibitem [{\citenamefont {Zaky}\ \emph {et~al.}(2020)\citenamefont {Zaky}, \citenamefont {Ahmed}, \citenamefont {Shalaby},\ and\ \citenamefont {Aly}}]{zaky2020refractive}%
  \BibitemOpen
  \bibfield  {author} {\bibinfo {author} {\bibfnamefont {Z.~A.}\ \bibnamefont {Zaky}}, \bibinfo {author} {\bibfnamefont {A.~M.}\ \bibnamefont {Ahmed}}, \bibinfo {author} {\bibfnamefont {A.~S.}\ \bibnamefont {Shalaby}},\ and\ \bibinfo {author} {\bibfnamefont {A.~H.}\ \bibnamefont {Aly}},\ }\bibfield  {title} {\bibinfo {title} {{Refractive index gas sensor based on the Tamm state in a one-dimensional photonic crystal: theoretical optimisation}},\ }\href@noop {} {\bibfield  {journal} {\bibinfo  {journal} {Scientific Reports}\ }\textbf {\bibinfo {volume} {10}},\ \bibinfo {pages} {1} (\bibinfo {year} {2020})}\BibitemShut {NoStop}%
\bibitem [{\citenamefont {Maji}\ and\ \citenamefont {Das}(2017)}]{maji2017hybrid}%
  \BibitemOpen
  \bibfield  {author} {\bibinfo {author} {\bibfnamefont {P.~S.}\ \bibnamefont {Maji}}\ and\ \bibinfo {author} {\bibfnamefont {R.}~\bibnamefont {Das}},\ }\bibfield  {title} {\bibinfo {title} {{Hybrid-Tamm-plasmon-polariton based self-reference temperature sensor}},\ }\href@noop {} {\bibfield  {journal} {\bibinfo  {journal} {Journal of Lightwave Technology}\ }\textbf {\bibinfo {volume} {35}},\ \bibinfo {pages} {2833} (\bibinfo {year} {2017})}\BibitemShut {NoStop}%
\bibitem [{\citenamefont {Buzavaite-Verteliene}\ \emph {et~al.}(2020)\citenamefont {Buzavaite-Verteliene}, \citenamefont {Plikusiene}, \citenamefont {Tolenis}, \citenamefont {Valavicius}, \citenamefont {Anulyte}, \citenamefont {Ramanavicius},\ and\ \citenamefont {Balevicius}}]{buzavaite2020hybrid}%
  \BibitemOpen
  \bibfield  {author} {\bibinfo {author} {\bibfnamefont {E.}~\bibnamefont {Buzavaite-Verteliene}}, \bibinfo {author} {\bibfnamefont {I.}~\bibnamefont {Plikusiene}}, \bibinfo {author} {\bibfnamefont {T.}~\bibnamefont {Tolenis}}, \bibinfo {author} {\bibfnamefont {A.}~\bibnamefont {Valavicius}}, \bibinfo {author} {\bibfnamefont {J.}~\bibnamefont {Anulyte}}, \bibinfo {author} {\bibfnamefont {A.}~\bibnamefont {Ramanavicius}},\ and\ \bibinfo {author} {\bibfnamefont {Z.}~\bibnamefont {Balevicius}},\ }\bibfield  {title} {\bibinfo {title} {{Hybrid Tamm-surface plasmon polariton mode for highly sensitive detection of protein interactions}},\ }\href@noop {} {\bibfield  {journal} {\bibinfo  {journal} {Optics Express}\ }\textbf {\bibinfo {volume} {28}},\ \bibinfo {pages} {29033} (\bibinfo {year} {2020})}\BibitemShut {NoStop}%
\bibitem [{\citenamefont {Augui{\'e}}\ \emph {et~al.}(2015)\citenamefont {Augui{\'e}}, \citenamefont {Bruchhausen},\ and\ \citenamefont {Fainstein}}]{auguie2015critical}%
  \BibitemOpen
  \bibfield  {author} {\bibinfo {author} {\bibfnamefont {B.}~\bibnamefont {Augui{\'e}}}, \bibinfo {author} {\bibfnamefont {A.}~\bibnamefont {Bruchhausen}},\ and\ \bibinfo {author} {\bibfnamefont {A.}~\bibnamefont {Fainstein}},\ }\bibfield  {title} {\bibinfo {title} {{Critical coupling to Tamm plasmons}},\ }\href@noop {} {\bibfield  {journal} {\bibinfo  {journal} {Journal of Optics}\ }\textbf {\bibinfo {volume} {17}},\ \bibinfo {pages} {035003} (\bibinfo {year} {2015})}\BibitemShut {NoStop}%
\bibitem [{\citenamefont {Wang}\ \emph {et~al.}(2024)\citenamefont {Wang}, \citenamefont {Li}, \citenamefont {Li}, \citenamefont {Gao}, \citenamefont {Yin}, \citenamefont {Liu}, \citenamefont {Zhong}, \citenamefont {Kan}, \citenamefont {Wang}, \citenamefont {Jiang},\ and\ \citenamefont {Shen}}]{Wang2024}%
  \BibitemOpen
  \bibfield  {author} {\bibinfo {author} {\bibfnamefont {Y.}~\bibnamefont {Wang}}, \bibinfo {author} {\bibfnamefont {Z.}~\bibnamefont {Li}}, \bibinfo {author} {\bibfnamefont {X.}~\bibnamefont {Li}}, \bibinfo {author} {\bibfnamefont {K.}~\bibnamefont {Gao}}, \bibinfo {author} {\bibfnamefont {Z.}~\bibnamefont {Yin}}, \bibinfo {author} {\bibfnamefont {W.}~\bibnamefont {Liu}}, \bibinfo {author} {\bibfnamefont {B.}~\bibnamefont {Zhong}}, \bibinfo {author} {\bibfnamefont {G.}~\bibnamefont {Kan}}, \bibinfo {author} {\bibfnamefont {X.}~\bibnamefont {Wang}}, \bibinfo {author} {\bibfnamefont {J.}~\bibnamefont {Jiang}},\ and\ \bibinfo {author} {\bibfnamefont {Z.}~\bibnamefont {Shen}},\ }\bibfield  {title} {\bibinfo {title} {Highly absorbing monolayer mos2 for a large reflection phase modulation},\ }\bibfield  {journal} {\bibinfo  {journal} {Advanced Optical Materials}\ }\textbf {\bibinfo {volume} {12}},\ \href {https://doi.org/10.1002/adom.202400429} {10.1002/adom.202400429} (\bibinfo {year} {2024})\BibitemShut
  {NoStop}%
\bibitem [{\citenamefont {Ferrier}\ \emph {et~al.}(2019)\citenamefont {Ferrier}, \citenamefont {Nguyen}, \citenamefont {Jamois}, \citenamefont {Berguiga}, \citenamefont {Symonds}, \citenamefont {Bellessa},\ and\ \citenamefont {Benyattou}}]{ferrier2019tamm}%
  \BibitemOpen
  \bibfield  {author} {\bibinfo {author} {\bibfnamefont {L.}~\bibnamefont {Ferrier}}, \bibinfo {author} {\bibfnamefont {H.~S.}\ \bibnamefont {Nguyen}}, \bibinfo {author} {\bibfnamefont {C.}~\bibnamefont {Jamois}}, \bibinfo {author} {\bibfnamefont {L.}~\bibnamefont {Berguiga}}, \bibinfo {author} {\bibfnamefont {C.}~\bibnamefont {Symonds}}, \bibinfo {author} {\bibfnamefont {J.}~\bibnamefont {Bellessa}},\ and\ \bibinfo {author} {\bibfnamefont {T.}~\bibnamefont {Benyattou}},\ }\bibfield  {title} {\bibinfo {title} {Tamm plasmon photonic crystals: from bandgap engineering to defect cavity},\ }\href@noop {} {\bibfield  {journal} {\bibinfo  {journal} {Apl Photonics}\ }\textbf {\bibinfo {volume} {4}},\ \bibinfo {pages} {106101} (\bibinfo {year} {2019})}\BibitemShut {NoStop}%
\bibitem [{\citenamefont {Gubaydullin}\ \emph {et~al.}(2017)\citenamefont {Gubaydullin}, \citenamefont {Symonds}, \citenamefont {Benoit}, \citenamefont {Ferrier}, \citenamefont {Benyattou}, \citenamefont {Jamois}, \citenamefont {Lema{\^\i}tre}, \citenamefont {Senellart}, \citenamefont {Kaliteevski},\ and\ \citenamefont {Bellessa}}]{gubaydullin2017tamm}%
  \BibitemOpen
  \bibfield  {author} {\bibinfo {author} {\bibfnamefont {A.}~\bibnamefont {Gubaydullin}}, \bibinfo {author} {\bibfnamefont {C.}~\bibnamefont {Symonds}}, \bibinfo {author} {\bibfnamefont {J.-M.}\ \bibnamefont {Benoit}}, \bibinfo {author} {\bibfnamefont {L.}~\bibnamefont {Ferrier}}, \bibinfo {author} {\bibfnamefont {T.}~\bibnamefont {Benyattou}}, \bibinfo {author} {\bibfnamefont {C.}~\bibnamefont {Jamois}}, \bibinfo {author} {\bibfnamefont {A.}~\bibnamefont {Lema{\^\i}tre}}, \bibinfo {author} {\bibfnamefont {P.}~\bibnamefont {Senellart}}, \bibinfo {author} {\bibfnamefont {M.}~\bibnamefont {Kaliteevski}},\ and\ \bibinfo {author} {\bibfnamefont {J.}~\bibnamefont {Bellessa}},\ }\bibfield  {title} {\bibinfo {title} {Tamm plasmon sub-wavelength structuration for loss reduction and resonance tuning},\ }\href@noop {} {\bibfield  {journal} {\bibinfo  {journal} {Applied Physics Letters}\ }\textbf {\bibinfo {volume} {111}},\ \bibinfo {pages} {261103} (\bibinfo {year} {2017})}\BibitemShut {NoStop}%
\bibitem [{\citenamefont {Girerd}\ \emph {et~al.}(2024)\citenamefont {Girerd}, \citenamefont {Mandorlo}, \citenamefont {Jamois}, \citenamefont {Benyattou}, \citenamefont {Ferrier},\ and\ \citenamefont {Berguiga}}]{Girerd2024}%
  \BibitemOpen
  \bibfield  {author} {\bibinfo {author} {\bibfnamefont {T.}~\bibnamefont {Girerd}}, \bibinfo {author} {\bibfnamefont {F.}~\bibnamefont {Mandorlo}}, \bibinfo {author} {\bibfnamefont {C.}~\bibnamefont {Jamois}}, \bibinfo {author} {\bibfnamefont {T.}~\bibnamefont {Benyattou}}, \bibinfo {author} {\bibfnamefont {L.}~\bibnamefont {Ferrier}},\ and\ \bibinfo {author} {\bibfnamefont {L.}~\bibnamefont {Berguiga}},\ }\bibfield  {title} {\bibinfo {title} {Optical sensing based on phase interrogation with a young’s interference hologram using a digital micromirror device},\ }\href {https://doi.org/10.1364/oe.507643} {\bibfield  {journal} {\bibinfo  {journal} {Optics Express}\ }\textbf {\bibinfo {volume} {32}},\ \bibinfo {pages} {3647} (\bibinfo {year} {2024})}\BibitemShut {NoStop}%
\bibitem [{\citenamefont {Qu}\ \emph {et~al.}(2015)\citenamefont {Qu}, \citenamefont {Ma}, \citenamefont {Hao}, \citenamefont {Qiu}, \citenamefont {Li}, \citenamefont {Xiao}, \citenamefont {Miao}, \citenamefont {Dai}, \citenamefont {He}, \citenamefont {Sun},\ and\ \citenamefont {Zhou}}]{Qu2015}%
  \BibitemOpen
  \bibfield  {author} {\bibinfo {author} {\bibfnamefont {C.}~\bibnamefont {Qu}}, \bibinfo {author} {\bibfnamefont {S.}~\bibnamefont {Ma}}, \bibinfo {author} {\bibfnamefont {J.}~\bibnamefont {Hao}}, \bibinfo {author} {\bibfnamefont {M.}~\bibnamefont {Qiu}}, \bibinfo {author} {\bibfnamefont {X.}~\bibnamefont {Li}}, \bibinfo {author} {\bibfnamefont {S.}~\bibnamefont {Xiao}}, \bibinfo {author} {\bibfnamefont {Z.}~\bibnamefont {Miao}}, \bibinfo {author} {\bibfnamefont {N.}~\bibnamefont {Dai}}, \bibinfo {author} {\bibfnamefont {Q.}~\bibnamefont {He}}, \bibinfo {author} {\bibfnamefont {S.}~\bibnamefont {Sun}},\ and\ \bibinfo {author} {\bibfnamefont {L.}~\bibnamefont {Zhou}},\ }\bibfield  {title} {\bibinfo {title} {Tailor the functionalities of metasurfaces based on a complete phase diagram},\ }\bibfield  {journal} {\bibinfo  {journal} {Physical Review Letters}\ }\textbf {\bibinfo {volume} {115}},\ \href {https://doi.org/10.1103/physrevlett.115.235503} {10.1103/physrevlett.115.235503} (\bibinfo {year}
  {2015})\BibitemShut {NoStop}%
\bibitem [{\citenamefont {Park}\ \emph {et~al.}(2016)\citenamefont {Park}, \citenamefont {Kang}, \citenamefont {Kim}, \citenamefont {Liu},\ and\ \citenamefont {Brongersma}}]{Park2016}%
  \BibitemOpen
  \bibfield  {author} {\bibinfo {author} {\bibfnamefont {J.}~\bibnamefont {Park}}, \bibinfo {author} {\bibfnamefont {J.-H.}\ \bibnamefont {Kang}}, \bibinfo {author} {\bibfnamefont {S.~J.}\ \bibnamefont {Kim}}, \bibinfo {author} {\bibfnamefont {X.}~\bibnamefont {Liu}},\ and\ \bibinfo {author} {\bibfnamefont {M.~L.}\ \bibnamefont {Brongersma}},\ }\bibfield  {title} {\bibinfo {title} {Dynamic reflection phase and polarization control in metasurfaces},\ }\href {https://doi.org/10.1021/acs.nanolett.6b04378} {\bibfield  {journal} {\bibinfo  {journal} {Nano Letters}\ }\textbf {\bibinfo {volume} {17}},\ \bibinfo {pages} {407–413} (\bibinfo {year} {2016})}\BibitemShut {NoStop}%
\bibitem [{\citenamefont {Zhou}\ \emph {et~al.}(2020)\citenamefont {Zhou}, \citenamefont {Li}, \citenamefont {Li}, \citenamefont {Yan}, \citenamefont {Zhang}, \citenamefont {Wang},\ and\ \citenamefont {Cheng}}]{zhou2020high}%
  \BibitemOpen
  \bibfield  {author} {\bibinfo {author} {\bibfnamefont {X.}~\bibnamefont {Zhou}}, \bibinfo {author} {\bibfnamefont {S.}~\bibnamefont {Li}}, \bibinfo {author} {\bibfnamefont {X.}~\bibnamefont {Li}}, \bibinfo {author} {\bibfnamefont {X.}~\bibnamefont {Yan}}, \bibinfo {author} {\bibfnamefont {X.}~\bibnamefont {Zhang}}, \bibinfo {author} {\bibfnamefont {F.}~\bibnamefont {Wang}},\ and\ \bibinfo {author} {\bibfnamefont {T.}~\bibnamefont {Cheng}},\ }\bibfield  {title} {\bibinfo {title} {{High-sensitivity SPR temperature sensor based on hollow-core fiber}},\ }\href@noop {} {\bibfield  {journal} {\bibinfo  {journal} {IEEE Transactions on Instrumentation and Measurement}\ }\textbf {\bibinfo {volume} {69}},\ \bibinfo {pages} {8494} (\bibinfo {year} {2020})}\BibitemShut {NoStop}%
\bibitem [{\citenamefont {Lu}\ \emph {et~al.}(2019)\citenamefont {Lu}, \citenamefont {Jiang}, \citenamefont {Hu}, \citenamefont {Zhou}, \citenamefont {Liu}, \citenamefont {Chen}, \citenamefont {Luo},\ and\ \citenamefont {Chen}}]{lu2019portable}%
  \BibitemOpen
  \bibfield  {author} {\bibinfo {author} {\bibfnamefont {L.}~\bibnamefont {Lu}}, \bibinfo {author} {\bibfnamefont {Z.}~\bibnamefont {Jiang}}, \bibinfo {author} {\bibfnamefont {Y.}~\bibnamefont {Hu}}, \bibinfo {author} {\bibfnamefont {H.}~\bibnamefont {Zhou}}, \bibinfo {author} {\bibfnamefont {G.}~\bibnamefont {Liu}}, \bibinfo {author} {\bibfnamefont {Y.}~\bibnamefont {Chen}}, \bibinfo {author} {\bibfnamefont {Y.}~\bibnamefont {Luo}},\ and\ \bibinfo {author} {\bibfnamefont {Z.}~\bibnamefont {Chen}},\ }\bibfield  {title} {\bibinfo {title} {{A portable optical fiber SPR temperature sensor based on a smart-phone}},\ }\href@noop {} {\bibfield  {journal} {\bibinfo  {journal} {Optics Express}\ }\textbf {\bibinfo {volume} {27}},\ \bibinfo {pages} {25420} (\bibinfo {year} {2019})}\BibitemShut {NoStop}%
\bibitem [{\citenamefont {Conteduca}\ \emph {et~al.}(2022)\citenamefont {Conteduca}, \citenamefont {Arruda}, \citenamefont {Barth}, \citenamefont {Wang}, \citenamefont {Krauss},\ and\ \citenamefont {Martins}}]{Conteduca2022}%
  \BibitemOpen
  \bibfield  {author} {\bibinfo {author} {\bibfnamefont {D.}~\bibnamefont {Conteduca}}, \bibinfo {author} {\bibfnamefont {G.~S.}\ \bibnamefont {Arruda}}, \bibinfo {author} {\bibfnamefont {I.}~\bibnamefont {Barth}}, \bibinfo {author} {\bibfnamefont {Y.}~\bibnamefont {Wang}}, \bibinfo {author} {\bibfnamefont {T.~F.}\ \bibnamefont {Krauss}},\ and\ \bibinfo {author} {\bibfnamefont {E.~R.}\ \bibnamefont {Martins}},\ }\bibfield  {title} {\bibinfo {title} {Beyond q: The importance of the resonance amplitude for photonic sensors},\ }\href {https://doi.org/10.1021/acsphotonics.2c00188} {\bibfield  {journal} {\bibinfo  {journal} {ACS Photonics}\ }\textbf {\bibinfo {volume} {9}},\ \bibinfo {pages} {1757–1763} (\bibinfo {year} {2022})}\BibitemShut {NoStop}%
\bibitem [{\citenamefont {Zhen}\ \emph {et~al.}(2014)\citenamefont {Zhen}, \citenamefont {Hsu}, \citenamefont {Lu}, \citenamefont {Stone},\ and\ \citenamefont {Soljačić}}]{Zhen2014}%
  \BibitemOpen
  \bibfield  {author} {\bibinfo {author} {\bibfnamefont {B.}~\bibnamefont {Zhen}}, \bibinfo {author} {\bibfnamefont {C.~W.}\ \bibnamefont {Hsu}}, \bibinfo {author} {\bibfnamefont {L.}~\bibnamefont {Lu}}, \bibinfo {author} {\bibfnamefont {A.~D.}\ \bibnamefont {Stone}},\ and\ \bibinfo {author} {\bibfnamefont {M.}~\bibnamefont {Soljačić}},\ }\bibfield  {title} {\bibinfo {title} {Topological nature of optical bound states in the continuum},\ }\bibfield  {journal} {\bibinfo  {journal} {Physical Review Letters}\ }\textbf {\bibinfo {volume} {113}},\ \href {https://doi.org/10.1103/physrevlett.113.257401} {10.1103/physrevlett.113.257401} (\bibinfo {year} {2014})\BibitemShut {NoStop}%
\end{thebibliography}%

\end{document}